\documentclass[%
 reprint,
superscriptaddress,
nofootinbib,
 amsmath,amssymb,
 aps,
]{revtex4-2}

\usepackage{graphicx}
\usepackage{dcolumn}
\usepackage{bm}
\usepackage{subcaption}
\usepackage{xcolor}
\usepackage{soul}

\providecommand{\abs}[1]{\lvert#1\rvert}
\providecommand{\bd}[1]{\boldsymbol{#1}}
\providecommand{\ro}[1]{\mathrm{#1}}

\providecommand{\Mp}{M_{\mathrm{Pl}}}

\usepackage[linktocpage]{hyperref}

\usepackage{hyperref}
\hypersetup{
    colorlinks,
    linkcolor={blue!100!black},
    citecolor={blue!100!black},
    urlcolor={blue!80!black}
}
\usepackage[mathlines]{lineno}

\begin{document}

\preprint{APS/123-QED}

\title{Gravothermalizing into primordial black holes, boson stars, and cannibal stars}

\author{Pranjal Ralegankar}%
 \email{pralegan@sissa.it}
\affiliation{
 SISSA, International School for Advanced Studies, via Bonomea 265, 34136 Trieste, Italy
}
\affiliation{
 INFN, Sezione di Trieste, via Valerio 2, 34127 Trieste, Italy
}
\affiliation{
 IFPU, Institute for Fundamental Physics of the Universe, via Beirut 2, 34014 Trieste, Italy
}

\author{Daniele Perri}%
 \email{dperri@sissa.it}
\affiliation{
 SISSA, International School for Advanced Studies, via Bonomea 265, 34136 Trieste, Italy
}
\affiliation{
 INFN, Sezione di Trieste, via Valerio 2, 34127 Trieste, Italy
}
\affiliation{
 IFPU, Institute for Fundamental Physics of the Universe, via Beirut 2, 34014 Trieste, Italy
}
\affiliation{
 Institute of Theoretical Physics, Faculty of Physics, University of
Warsaw, ul. Pasteura 5, PL-02-093 Warsaw, Poland
}

\author{Takeshi Kobayashi}%
 \email{takeshi.kobayashi@sissa.it}
\affiliation{
 SISSA, International School for Advanced Studies, via Bonomea 265, 34136 Trieste, Italy
}
\affiliation{
 INFN, Sezione di Trieste, via Valerio 2, 34127 Trieste, Italy
}
\affiliation{
 IFPU, Institute for Fundamental Physics of the Universe, via Beirut 2, 34014 Trieste, Italy
}


\begin{abstract}
Very little is known about the cosmological history from after the end of inflation until Big Bang Nucleosynthesis. Various well-motivated models predict that the universe could have undergone a period of matter domination in this early epoch.
We demonstrate that if the particles causing matter domination have self-interactions, they can form halos that undergo a gravothermal collapse. 
We thus propose a novel scenario for the formation of primordial black holes, which in particular can lie within the asteroid-mass range.
We also find that it is not only black holes that can form in the aftermath of a gravothermal evolution.
We show that number-changing annihilations of the particles
can create sufficient heat to halt the gravothermal evolution, thus forming a ``cannibal star''. Likewise, the pressure from the particle's repulsive self-interactions can form a boson star during a gravothermal evolution. 
These stars can also eventually collapse into black holes.
Thus, our study highlights that structure formation in the early universe can have a rich phenomenology.
\end{abstract}

\maketitle

\tableofcontents

\section{\label{sec:level1}Introduction}

Recent advances in precision cosmology have provided detailed insights into our universe’s history, from the earliest moments during the inflationary epoch to the periods following Big Bang Nucleosynthesis (BBN). However, the intermediate era---spanning potentially up to 36 orders of magnitude in time---remains largely unexplored.

One intriguing possibility is that the universe underwent an early matter-dominated era (EMDE) before BBN, see Ref.~\cite{Allahverdi:2020bys} for review. Such a phase can arise naturally from the coherent oscillation of the inflaton or a moduli field in a quadratic potential \cite{Turner:1983he,Kane:2015jia}. An EMDE is also common in scenarios involving hidden sectors that are thermally decoupled from the Standard Model (SM) particles \cite{Zhang:2015era, Berlin:2016gtr,Erickcek:2020wzd,Erickcek:2021fsu}. If the lightest particle in the hidden sector becomes non-relativistic, it behaves as pressureless matter that can drive an EMDE. 
The Hot Big Bang era begins when the particles causing EMDE decay into SM particles, which must occur before BBN.

If the EMDE lasts longer than 12 $e$-folds, then gravity has enough time to cause density perturbations to grow non-linear, leading to the formation of halos in the early universe \cite{Barenboim:2013gya,Jedamzik:2010dq,Erickcek:2011us,Fan:2014zua,Eggemeier:2021smj,Lozanov:2022yoy, Lozanov:2023knf,Padilla:2021zgm,Fernandez:2023ddy, Ganjoo:2023fgg}. Halos are prone to further collapse because self-gravitating systems in virial equilibrium have negative heat capacity: losing energy raises their temperature, which accelerates further energy loss. 
As higher temperature, or virial speed, implies that the system is more compact, halos are predisposed to become more compact as long as there is a mechanism for heat loss. 
In conventional astrophysical halos, energy loss typically occurs through radiative cooling. In EMDE scenarios without radiative channels, it may seem that compact object formation is suppressed. (See Refs.~\cite{Bramante:2024pyc,Savastano:2019zpr,Flores:2020drq} for dark compact object formation with radiative cooling.)

In this study, we highlight that if the particles causing EMDE have self-interactions, then that alone is sufficient to form compact objects, including primordial black holes (PBHs). Specifically, self-scatterings cause heat loss through particle ejections, triggering a process known as gravothermal collapse.

The gravothermal phenomenon was first studied for globular clusters, where the
gravitational self-interaction between stars is sufficient to transfer heat and induce gravothermal collapse \cite{1987degc.book,Lynden-Bell:1980xip}. In globular clusters, the gravothermal evolution is eventually stabilized by the formation of binaries. 
On the other hand, for collisionless dark matter halos, gravothermal collapse does not occur because its time scale 
grows with the number of dark matter particles and thus 
significantly exceeds the age of the universe.

In contrast, dark matter with non-gravitational self-interactions allows for a faster energy loss, enabling gravothermal collapse on shorter time scales \cite{Balberg:2001qg, Balberg:2002ue, Koda:2011yb, Pollack:2014rja, Essig:2018pzq, Feng:2020kxv, Feng:2021rst, Zeng:2021ldo, Xiao:2021ftk, Outmezguine:2022bhq}. The gravothermal process has also been shown to facilitate the transport of angular momentum from the collapsing core \cite{1976ApJ...210..757S,Feng:2021rst}.
It has been argued through both numerical and analytical techniques that black holes can be formed at the end of a gravothermal evolution in the case of self-interacting dark matter \cite{Balberg:2001qg, Feng:2021rst, Gad-Nasr:2023gvf}. However, current observational constraints on dark matter self-interactions limit the possibility of a collapse occurring by the present day for most halos \cite{Elbert:2014bma, Meshveliani:2022rih}.

The period of matter domination in the standard Hot Big Bang history lasts about eight~$e$-folds. An EMDE, on the other hand, can have a much longer duration. 
We demonstrate that if the particles responsible for the EMDE 
(but are not the dark matter in the present universe)
possess self-interactions, gravothermal collapse can readily occur before the Hot Big Bang era begins.

We also show that the gravothermal phenomenon can produce exotic compact objects other than PBHs. Here, it is important to note that 
the same self-interaction that enables a gravothermal evolution also gives rise to number-changing annihilations. These cannibal annihilations can generate enough heat to balance the gravothermal cooling, giving rise to compact objects which we refer to as ``cannibal stars''. Similarly, at large densities, the pressure from the particle's repulsive self-interactions can balance gravity and hence produce the so-called boson stars \cite{Hu:2000ke,Colpi:1986ye}. 
These novel outcomes broaden the range of possible structures that could arise in the early universe, and demonstrate that even simple particle models can give rise to complex astrophysical phenomena through the gravothermal catastrophe.

In this work, we aim to present a new paradigm for producing PBHs and exotic stars. Hence, for the EMDE model, we restrict ourselves to a simple case of a real scalar particle with a quartic self-interaction. 
For this toy model, we find that gravothermalized halos predominantly form cannibal stars.
The formation of PBHs requires the cannibal stars to accrete the surrounding particles and collapse.

Note that the gravothermal mechanism for generating PBHs is distinct from the popularly discussed scenario that invokes a direct collapse of large-amplitude curvature perturbations \cite{Carr:1974nx, ShamsEsHaghi:2022azq, Khlopov:2008qy}. Importantly, the gravothermal mechanism does not require an enhancement of inflationary perturbations and hence avoids the fine-tuning issues faced by 
the direct collapse scenario \cite{Cole:2023wyx}.

This paper is organized as follows. In section~\ref{sec:emde}, we outline our EMDE model. In section~\ref{sec:grav}, we discuss the formation and gravothermal evolution of halos during an EMDE. 
In section~\ref{sec:times}, we estimate the masses of PBHs that can be produced. We also evaluate the conditions for cannibal or boson stars to be produced instead of black holes. In section~\ref{sec:BH}, we provide an estimate of the PBH abundance. In section~\ref{sec:param}, we illustrate the EMDE parameter space that can produce PBHs. In section~\ref{sec:beyond}, we comment on alternative EMDE scenarios. Finally, we conclude in section~\ref{sec:end}. Some of the calculations are relegated to appendices. In appendix~\ref{app:perturb}, we review the evolution of density perturbations during matter domination. In appendix~\ref{app:press}, we review the Press--Schechter formalism, which we use to estimate the PBH abundance.

We work in natural units where the speed of light, the reduced Planck's constant, and the Boltzmann constant are set to unity. We use $M_{\rm Pl} = (8 \pi G)^{-1/2} =2.435\times 10^{18}\, \ro{GeV}$ to denote the reduced Planck mass.

\section{EMDE model}\label{sec:emde}
There are several ways an EMDE could begin. For simplicity, in this study, we largely focus on a scenario where the EMDE begins once the particles causing the EMDE, $\phi$, become non-relativistic. Specifically, we consider that the universe is initially dominated by $\phi$ particles and that $\phi$ particles are thermally distributed. At early enough periods, the temperature of the $\phi$ particles is much larger than their mass, $m$, and hence, the universe is radiation-dominated. Once the expansion of the universe cools $\phi$ particles to non-relativistic energies, the universe enters the EMDE. 

We denote the scale factor at the beginning of the EMDE by~$a_{\mathrm{i}}$.
This corresponds to the time when the evolution of $\phi$'s energy density transitions from $\rho_{\phi}\propto 1/a^4$ to $\rho_{\phi}\propto 1/a^3$. The transition occurs roughly when the temperature of $\phi$ becomes of order $m/3$; it was shown numerically in Ref.~\cite{Ganjoo:2022rhk} 
that the energy density of $\phi$ deep in the EMDE is well approximated by $\rho_\phi (a_{\mathrm{i}}) (a_{\mathrm{i}} / a)^3$ with
\begin{align}\label{eq:rho_ai}
    \rho_{\phi}(a_{\mathrm{i}})\approx \frac{\pi^2}{30}\left(\frac{m}{3}\right)^4,
\end{align}
given that $\phi$ is a scalar particle.
Thus, we use this expression as the energy density of $\phi$ (and the universe) at the beginning of the EMDE.

The EMDE ends when $\phi$ decays into relativistic SM particles. As in previous literature, we call the end of the EMDE as reheating. Considering $a_{\rm rh}$ to be the scale factor at reheating and $T_{\rm rh}$ to be the temperature of the SM plasma at that time, the energy of the universe at the end of EMDE is given by 
\begin{align}
 \rho_{\mathrm{SM}}(a_{\mathrm{rh}})= \frac{\pi^2}{30}g_{*\mathrm{rh}}T_{\mathrm{rh}}^4.
\label{eq:rho_rh}
\end{align}
Here, $g_{* \mathrm{rh}}$ is the number of effective degrees of freedom in the SM plasma at $T_{\mathrm{rh}}$. Since we are mostly interested in order-of-magnitude estimates, for brevity, we set $g_{*  \mathrm{rh}} = g_{* s, \mathrm{rh}} =100$ in this paper. We have checked that most of our results do not change significantly if $g_{* \mathrm{rh}}$ changes by an order of magnitude.

One can obtain a direct relationship between $m$ and $T_{\rm rh}$ by using the fact that almost all of the energy from $\phi$ particles goes to SM particles at $a_{\rm rh}$,
\begin{align}
 \rho_{\mathrm{SM}}(a_{\mathrm{rh}})\approx \rho_{\phi}(a_{\mathrm{i}})\left(\frac{a_{\mathrm{i}}}{a_{\mathrm{rh}}}\right)^{3}.   
\end{align}
Using the expressions for $\rho_{\mathrm{SM}}(a_{\mathrm{rh}})$ and $\rho_{\phi}(a_{\mathrm{i}})$, 
we obtain
 \begin{align}\label{eq:mass}
    m\approx 9 \, T_{\mathrm{rh}} \left( \frac{a_{\mathrm{rh}}}{a_{\mathrm{i}}} \right)^{3/4}.
\end{align}

Halos that form during the EMDE can undergo gravothermal evolution, given that the $\phi$~particles have self-interactions. 
As a toy model, we consider a real scalar with a quartic self-interaction,
\begin{align}
    V=\frac{1}{2}m^2\phi^2+\frac{\lambda}{4!}\phi^4.
\label{eq:phi4}
\end{align}
Unless otherwise specified, we take $\lambda$ to be positive.

Our EMDE model has three independent parameters. In this study, we choose these parameters to be the scale of reheating~$T_{\mathrm{rh}}$, the duration of EMDE~$a_{\mathrm{rh}}/a_{\mathrm{i}}$, and the self-interaction coupling~$\lambda$ of the particle. The mass~$m$ is  given in terms of $T_{\mathrm{rh}}$ and $a_{\mathrm{rh}}/a_{\mathrm{i}}$
via eq.~\eqref{eq:mass}.

For other EMDE models, such as those in which non-relativistic particles dominate over radiation, or a scalar field that oscillates in a $\phi^2$ potential, the primary 
difference arises in the relation between the mass~$m$ and the initial density~$\rho_{\phi}(a_{\ro{i}})$. 
Apart from this point, our results can also largely be applied to these other EMDE models. We discuss this in section~\ref{sec:beyond}.

\section{Gravothermal evolution during an EMDE}\label{sec:grav}

In this section, we first review the growth of perturbations in a matter-dominated era and the formation of halos as the perturbations become non-linear. 
We then outline the relevant timescale of the gravothermal evolution during EMDE and discuss how halo cores form and eventually collapse. 

\subsection{Halo formation}
\label{sec:halo}

A notable feature of an era dominated by pressureless matter is the growth of matter density perturbations~$\delta$ due to gravity.
To be specific, let us consider Fourier modes with wave number~$k$ that enter the horizon during EMDE, i.e.,
\begin{align}\label{eq:klight}
  (aH)_{\rm rh}<k<(aH)_{\mathrm{i}}.
\end{align}
When these modes are deep inside the horizon, the dimensionless power spectrum of the linear density perturbation grows as (see appendix~\ref{app:perturb} for derivation)
\begin{align}\label{eq:delta_ev} \Delta^2_\delta(k,a)=
\frac{4}{25}
\left(\frac{a}{a_{\mathrm{hor}}} \right)^2
\Delta^2_\mathcal{R}(k).
\end{align}
Here $\Delta^2_\mathcal{R} (k)$ is the dimensionless curvature power spectrum when the mode was outside the horizon, and $a_{\mathrm{hor}}$ is the scale factor at horizon entry, i.e.,
\begin{align}
    (aH)_{\mathrm{hor}}\equiv k.
\label{eq:eight}
\end{align}
As $H\propto a^{-3/2}$ during EMDE, and denoting the wave mode that enters the horizon at the end of the EMDE by 
\begin{align}\label{eq:krh}
    k_{\rm rh}\equiv  (aH)_{\rm rh}, 
\end{align}
the scale factor at horizon entry can also be written as
\begin{align}
a_{\mathrm{hor}}=a_{\mathrm{rh}}\left(\frac{k_{\mathrm{rh}}}{k}\right)^2.
\label{eq:a_hor}
\end{align}

For modes entering the horizon prior to $a_{\mathrm{i}}$, i.e. $k>(aH)_{\mathrm{i}}$, the density perturbations initially experience the relativistic nature of the $\phi$ particles. If the self-interaction is strong enough to maintain kinetic equilibrium, then the density perturbations undergo acoustic oscillations even beyond $a_{\ro{i}}$ and thus experience less growth compared to those with larger wavelengths \cite{Ganjoo:2022rhk}. If instead the self-interactions are negligible, then the free-streaming of $\phi$ suppresses density perturbations. In either case, the matter power spectrum is suppressed for $k>(aH)_{\mathrm{i}}$, and thus we ignore such modes.

Once the density contrast becomes of order unity, the matter perturbations become non-linear, and eq.~\eqref{eq:delta_ev} breaks down. In this non-linear regime, the particles in overdense regions collapse to form a halo.
We define $a_{\ro{NL}}$ as the scale factor when the linear density contrast becomes
\begin{align}
   \Delta_\delta(k, a_{\rm NL})=1.686,
\label{eq:a_NL}
\end{align}
with the left-hand side calculated using eq.~\eqref{eq:delta_ev}. 
This can be considered as the moment when halos corresponding to the wave number~$k$ form.\footnote{More precisely, $a_{\ro{NL}}(k)$ defined via eq.~(\ref{eq:a_NL}) corresponds to the time when the number of halos with mass $M_{\ro{halo}}(k)$ (cf. eq.~(\ref{eq:mhalo1})) is around its maximum. See Appendix~\ref{app:press}.}
Here we remark that the stochasticity in density perturbations induces halos of a given~$k$ to collapse at different times, and $a_{\rm NL}(k)$ defined above denotes the mean of the collapse scale factors.
In this study, we are primarily interested in order-of-magnitude estimates of the PBH abundance and hence we ignore the stochasticity.
The dimensionless curvature power spectrum~$\Delta^2_{\mathcal{R}} (k)$ is roughly scale-invariant and is of order $\sim 10^{-9}$ \cite{Planck:2018jri}. Hence we find\footnote{Most results in this paper have a precision only at the order of magnitude level, as we indicate by~$\sim$. However, we report one significant digit in the intermediate steps to avoid errors from being magnified by large powers in the final results.}
\begin{align}\label{eq:acoll}
    a_{\ro{NL}}\sim 1\times 10^5a_{\mathrm{hor}}=1\times 10^5a_{\mathrm{rh}}\left(\frac{k_{\mathrm{rh}}}{k}\right)^2.
\end{align}

The halo mass is determined by the wave number of the perturbation mode that collapses.
Specifically, matter in a radius of 
$r\sim a_{\rm NL}/k$ collapses to form a halo of mass 
\begin{align}\label{eq:mhalo1}
    M_{\rm halo}= \frac{4\pi}{3}\rho_{\phi}(a_{\rm NL})\left(\frac{a_{\rm NL}}{k}\right)^3.    
\end{align}
We can rewrite $M_{\rm halo}$ in terms of the EMDE parameters by noting that $\rho_{\phi}(a_{\rm NL})a_{\rm NL}^3=\rho_{\ro{SM}}(a_{\mathrm{rh}})a_{\mathrm{rh}}^3$.
Further using eqs.~\eqref{eq:rho_rh}, \eqref{eq:krh}, 
and $\rho_{\ro{SM}}(a_{\mathrm{rh}}) = 3 \Mp^2 H_{\ro{rh}}^2$, one obtains,
\begin{align}\label{eq:mhalo}
    M_{\rm halo}\approx 1\times10^{22}\ {\rm g} \ \left(\frac{T_{\mathrm{rh}}}{\rm 100 \, MeV}\right)^{-2}\left(\frac{k/k_{\mathrm{rh}}}{10^4}\right)^{-3}.
\end{align}
Given that halos form only from $k$ modes for which $a_{\rm NL}<a_{\rm rh}$, and also requiring $T_{\rm rh} \gtrsim 10$~MeV, the halo mass during the EMDE is bounded as
$M_{\ro{halo}} \lesssim 10^{28}\ {\rm g}$.

The self-interaction of the particles plays a negligible role during the growth of linear density perturbations as well as in the determination of $a_{\rm NL}$ or $M_{\rm halo}$. However, it can become important after perturbations become non-linear, i.e. during halo formation. 
Let us here focus on cases where self-interactions are negligible at the initial stage of the halo formation. (We later present the condition under which this assumption is valid.) 
Then, the halo is initially collisionless and well described by the NFW profile \cite{Navarro:1995iw},
\begin{equation}
\rho_{\ro{NFW}}(r)=\frac{\rho_s}{\frac{r}{r_s}\left(1+\frac{r}{r_s}\right)^2},
\label{eq:NFW}
\end{equation}
where $\rho_s$ and $r_s$ are the scale density and scale radius, respectively.

The parameters of the NFW profile can be connected to those of the background cosmology, by 
identifying $M_{\ro{halo}}$ in~eq.~\eqref{eq:mhalo1} with the virial mass, 
and also assuming that within the virial radius 
(i.e. radius enclosing~$M_{\ro{halo}}$), the mean density 
is $200\rho_{\phi}(a_{\rm NL})$. We define the halo radius as the virial radius of the halo of mass $M_{\rm halo}(k)$ at $a_{\rm NL}(k)$, i.e.
\begin{align}\label{eq:rhalo}
    r_{\rm halo}(k)=200^{-1/3}\frac{a_{\rm NL}(k)}{k}.
\end{align}
Its ratio to the scale radius defines the concentration at $a_{\rm NL}$,
\begin{align}
    c=\frac{r_{\rm halo}}{r_s}.
\end{align}
In the literature, the virial radius is defined such that it increases with time (proportional to the scale factor). Consequently, the concentration factor evolves with time even if the halo central profile is stationary \cite{Diemer:2018vmz}. However, we stress that in this paper, we use $c$ to denote the concentration at $a_{\rm NL}$.

As the mass of the halo is the mass contained within $r_{\rm halo}$, for the NFW profile we obtain
\begin{align}
    M_{\rm halo}=4\pi\rho_sr_s^3h(c),
\end{align}
where $h(c)=\log(1+c)-c/(1+c)$.
The scale and background densities can be related by 
replacing $M_{\rm halo}$ using eq.~\eqref{eq:mhalo1}, then using eq.~\eqref{eq:rhalo} to replace $k$, and finally using $r_{\rm halo}=cr_s$. This yields,
\begin{align}\label{eq:conc_relations}
    \rho_s=&\frac{200\rho_{\phi}(a_{\rm NL})}{3}\frac{c^3}{h(c)}.
\end{align}
Thus, one finds that halos corresponding to larger wave numbers have larger central densities because these modes collapse earlier when $\rho_{\phi}$ is larger.

We define the virial velocity of a halo as
\begin{align}
    v_{\rm vir}^2\equiv \frac{GM_{\rm halo}}{r_{\rm halo} }.
\label{eq:vvir}
\end{align}
We also define the scale velocity as $v_s\equiv r_s\sqrt{4\pi G\rho_s}$.
These are related as,
\begin{align}\label{eq:vs}
    v_{s}=~&v_{\rm vir}\sqrt{\frac{c}{h(c)}}.
\end{align}
One can find the value of the virial velocity by replacing $M_{\rm halo}$ and $r_{\rm halo}$ via eqs.~\eqref{eq:mhalo1} and \eqref{eq:rhalo}. Further using eq.~(\ref{eq:eight}) and 
$\rho_{\phi} (a_{\ro{NL}}) = 3 \Mp^2 H_{\ro{NL}}^2$, we obtain
\begin{align}
    v_{\mathrm{vir}}^2 =  \frac{200^{1/3}}{2} \frac{(aH)^2_{\mathrm{NL}}}{(aH)_{\mathrm{hor}}^2}=
\frac{200^{1/3}}{2} \frac{a_{\mathrm{hor}}}{a_{\rm NL}}.
\end{align}
In the second relation, we used the fact that the universe is matter dominated between the horizon entry of the mode and halo formation.
This, combined with (\ref{eq:acoll}), yields (see also \cite{Blanco:2019eij})
\begin{align}
    v_{\mathrm{vir}}\sim 0.5\times 10^{-2}.
\label{eq:v_vir}
\end{align}

To connect the halo properties with the background cosmology one needs to numerically calculate $c$ through simulations. Ref.~\cite{Zhao:2008wd} has performed N-body simulations for a matter-dominated universe with a power spectrum taking a form 
$P(k) = (2 \pi^2/ k^3) \Delta_\delta^2(k) \propto k$,
which corresponds to the EMDE universe under consideration
(cf. eqs. (\ref{eq:delta_ev}) and (\ref{eq:a_hor})). 
According to their fig.~19, the concentration of halos at their formation is $c\approx 30$.\footnote{Fig.~19 in Ref.~\cite{Zhao:2008wd} shows concentration as a function of the halo mass normalized by the mass when it becomes non-linear. Thus, $c$ can be read off from where the normalized mass is unity.}
Since we are ignoring the stochasticity in halo properties, in the rest of the paper we fix all halos to have $c=30$.
In particular, with this value for~$c$, one can relate the scale density 
at halo formation
in eq.~(\ref{eq:conc_relations}) with the 
reheating scales using eq.~(\ref{eq:acoll}) as,
\begin{equation}
\rho_s \sim 1 \times 10^{16}\, T_{\ro{rh}}^4
\left( \frac{k/k_{\ro{rh}}}{10^4} \right)^6.
\label{eq:rho_s}
\end{equation}

\subsection{Gravothermal time scale}

After the halo is formed, self-scatterings cause a redistribution of the $\phi$~particles and give rise to a uniform-density core at the center of the halo. 
Considering that the influences of other halos (such as their tidal force) are negligible,
the redistribution occurs on a relaxation time scale determined by the
mean time between the particle collisions, $\tau_r\sim1/(\rho v \sigma)$
\cite{Balberg:2002ue,Pollack:2014rja,Outmezguine:2022bhq,Koda:2011yb,Essig:2018pzq}.
Here, $\sigma$ is the self-scattering cross section per unit mass,
which for the model in~eq.~\eqref{eq:phi4} is
\begin{equation}\label{eq:sigma}
 \sigma=\frac{\lambda^2}{128 \pi m^3},
\end{equation}
at tree level and in the non-relativistic limit. For an NFW profile, let us write the relaxation time scale in terms of the scale density and velocity as
\begin{align}\label{eq:tr}
    \tau_{r}^{\rm NFW}=\frac{1}{\rho_s v_{s} \sigma }.
\end{align}
Note that our definition of the relaxation time scale is about a factor of two different from those in Refs.~\cite{Balberg:2001qg,Koda:2011yb}.

Since we focus on halos that are initially almost collisionless (cf. discussions above eq.~(\ref{eq:NFW})), the relaxation time scale needs to be larger than the dynamical time scale of the halo, $\tau_{\rm dyn}=1/\sqrt{4\pi G\rho}$. 
For the NFW profile, let us rewrite this using the scale density as\footnote{The NFW dynamical time is much shorter than the Hubble time upon halo formation: 
$\tau_{\rm dyn}^{\rm NFW} \sim 10^{-3} / H_{\ro{NL}} $ for $c = 30$.}
\begin{align}\label{eq:tdyn}
    \tau_{\rm dyn}^{\rm NFW}=\frac{1}{\sqrt{4\pi G\rho_s}}.
\end{align}
Ref.~\cite{Essig:2018pzq} explicitly demonstrated that an NFW halo can initially be treated as collisionless, i.e. optically thin, if 
\begin{equation}
 \tau_{\rm dyn}^{\rm NFW} < \tau_{r}^{\rm NFW}.
\label{eq:thin}
\end{equation}
This situation is also known as the long mean free path (LMFP) regime, as the mean free path of the particles is larger than the scale radius of the halo. As $\tau_{\rm dyn}^{\rm NFW}/\tau_{r}^{\rm NFW}\propto v_s \sqrt{\rho_s}$, and $\rho_s$ increases with $k$
while $v_s$ is $k$-independent (see eqs.~(\ref{eq:conc_relations}) and (\ref{eq:vs})), 
the requirement of~eq.~\eqref{eq:thin} imposes an upper bound on $k$,
\begin{align}\label{eq:knfw}
    \frac{k}{k_{\rm rh}} \lesssim 6 \times 10^4\lambda^{-2/3}\left(\frac{T_{\mathrm{rh}}}{\rm 100~ MeV}\right)^{1/3}\left( \frac{a_{\mathrm{rh}}/a_{\mathrm{i}}}{10^{10}} \right)^{3/4}.
\end{align}
In this study, we typically focus on modes satisfying this condition.

After the formation of the core, gravothermal evolution ensues, where the central core density keeps increasing on a relaxation time scale. 
This relaxation time becomes shorter with the increase in the central density.
Consequently, the core undergoes a gravothermal collapse within a finite time after halo formation \cite{Lynden-Bell:1980xip}. 
Numerical studies~\cite{Koda:2011yb,Outmezguine:2022bhq,Balberg:2001qg,Essig:2018pzq} have found that the time interval for an initially NFW halo to collapse is $\sim 200\tau_r^{\rm NFW}$. 
We can thus express the time when gravothermal collapse happens as
\begin{align}
 t_{\rm GC} \sim t_{\rm NL} + 200\tau_{r}^{\rm NFW}.
\label{eq:t_GC}
\end{align}

As the rate of gravothermal evolution becomes faster with time, most of the increase in $\rho_c$ occurs near $t_{\rm GC}$. Consequently, the formation of a final stable compact object (e.g. a black hole) is expected to take place also at $\sim t_{\rm GC}$.\footnote{We stress that this discussion assumes the halo to remain isolated from external tidal forces until it collapses. Since most halos undergo continual mergers and accretion, it is only rare halos that can stay isolated for significant periods; we estimate the abundance of such halos in Section~\ref{sec:BH}.}

Considering that the gravothermal evolution takes place during EMDE, 
eq.~(\ref{eq:t_GC}) can be rewritten in terms of the Hubble rates at halo formation and collapse as
\begin{align}\label{eq:agc_def}
   \frac{2}{3}\left(\frac{1}{H_{\rm GC}}-\frac{1}{H_{\rm NL}}\right) \sim  200\tau_{r}^{\rm NFW}.
\end{align} 
This can be solved to obtain the scale factor upon gravothermal collapse. 
If $200\tau_{r}^{\rm NFW}\ll 1/H_{\rm NL}$, then $a_{\rm GC}$ is almost the same as $a_{\rm NL}$. In contrast if $200\tau_{r}^{\rm NFW}\gg 1/H_{\rm NL}$, then one can solve for $a_{\rm GC}$ by noting that
$H_{\rm GC}=H_{\mathrm{rh}}(a_{\mathrm{rh}}/a_{\rm GC})^{3/2}$.
We also remark that the relaxation time is determined 
via eq.~\eqref{eq:tr} by the scales of the initial NFW profile, 
which in turn are functions of the EMDE parameters through 
eqs.~(\ref{eq:vs}) and (\ref{eq:rho_s}).
Hence one finds,
\begin{align}\label{eq:abh}
    \frac{a_{\rm GC}}{a_{\rm i}}\sim 
\max\bigg[ & 2\times 10^{8} \, \lambda^{-4/3} \left(\frac{k/k_{\mathrm{rh}}}{10^4}\right)^{-4}\left(\frac{T_{\mathrm{rh}}}{\rm 100\, MeV}\right)^{2/3}
\nonumber\\
&\times\left( \frac{a_{\mathrm{rh}}/a_{\mathrm{i}}}{10^{10}} \right)^{5/2}, \ \ \frac{a_{\rm NL}}{a_{\rm i}}\bigg].
\end{align}
The first (second) term in the square brackets gives $a_{\ro{GC}} / a_{\rm i}$
for $200\tau_{r}^{\rm NFW} \gg (\ll) 1/H_{\rm NL}$.

The above expression for $a_{\rm GC}$ is valid only if gravothermal collapse occurs before reheating, i.e.,
\begin{align}
    a_{\rm GC}<a_{\rm rh}.
\label{eq:33}
\end{align}
This requirement translates into a lower bound on $k$,
\begin{align}\label{eq:kheavy}
    \frac{k}{k_{\rm rh}} \gtrsim \max\bigg[
&4\times 10^3\ \lambda^{-1/3}\left(\frac{T_{\mathrm{rh}}}{\rm 100~MeV}\right)^{1/6}\nonumber\\
    &\times \left( \frac{a_{\mathrm{rh}}/a_{\mathrm{i}}}{10^{10}} \right)^{3/8},\ \ 4 \times 10^{2}\bigg].
\end{align}
This sets a necessary condition for a given $k$~mode to undergo a gravothermal collapse during the EMDE. 

We also note that for initially collisionless halos,
$200\tau_{r}^{\rm NFW} $ is comparable to or 
larger than $1/H_{\rm NL}$:
\begin{equation}
 200\tau_{r}^{\rm NFW} H_{\rm NL} 
\sim 0.2 \, 
\frac{\tau_{r}^{\rm NFW}}{\tau_{\rm dyn}^{\rm NFW}}
> 0.2.
\label{eq:marg}
\end{equation}
The first relation follows from eqs.~\eqref{eq:conc_relations} and \eqref{eq:tdyn}, while the inequality corresponds to the collisionless condition in eq.~(\ref{eq:thin}).
Hence, for initially collisionless halos, eqs.~(\ref{eq:abh}) and (\ref{eq:kheavy}) are set by the first terms in the square brackets.

\begin{figure}
    \includegraphics[width=0.49\textwidth]{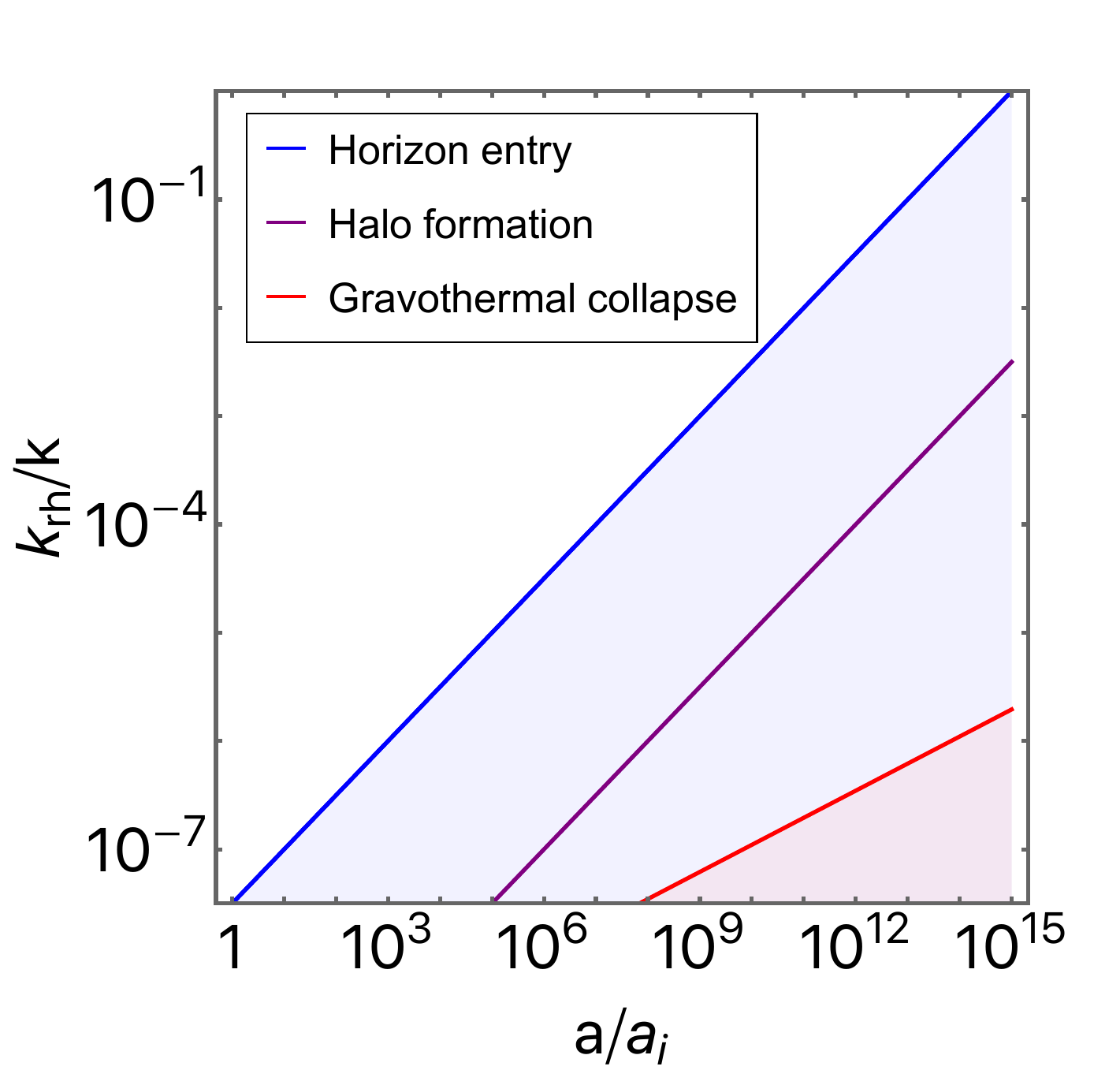}
    \caption{\label{fig:scales} 
Scale factor during EMDE when perturbation modes with fixed comoving length scales ($k^{-1}$) enter the horizon (blue line), become non-linear and form halos (purple), and when the corresponding halos undergo gravothermal collapse (red). Here we use $\lambda = 10^{-1}$, $T_{\mathrm{rh}} = 100\, \mathrm{MeV}$, and $a_{\mathrm{rh}}/a_{\mathrm{i}} = 10^{15}$.}
\end{figure}

In Fig.~\ref{fig:scales}, we illustrate when perturbation modes with fixed comoving wavelengths enter the horizon (blue), become non-linear (purple), and undergo a gravothermal collapse (red). 
In other words, the lines show respectively $a_{\mathrm{hor}}$, $a_{\rm NL}$, and $a_{\rm GC}$ as functions of $k^{-1}$.
The EMDE parameters are chosen here such that the lines are well separated.
One sees that for larger wavelengths, the separation $a_{\ro{GC}} / a_{\ro{NL}}$ between gravothermal collapse and halo formation increases.
This is a consequence of the fact that longer-wavelength modes form halos later and with lower central densities, which in turn lead to slower gravothermal evolutions due to 
$\tau_r\propto 1/\rho$. 

From the above discussion, one may naively expect gravothermal effects to be more prominent if the EMDE begins earlier, namely, at a larger~$\rho_\phi (a_{\ro{i}})$.
However, this is not necessarily the case, since the
self-interaction cross section can also be related to the cosmological density.
To be precise, let us evaluate the `efficiency' of the gravothermal evolution by comparing the relaxation time with the dynamical time, 
\begin{align}
    \frac{\tau_r^{\rm NFW}}{\tau^{\rm NFW}_{\rm dyn}}\propto 
\frac{1}{\sigma \sqrt{\rho_{\phi}(a_{\rm NL})} }\propto 
[\rho_\phi (a_{\ro{i}})]^{1/4}
 \left( \frac{k}{k_{\ro{i}}} \right)^{3}.
\end{align}
Here, we have focused on the dependence on the background density.
Upon moving to the far right-hand side, we introduced 
$k_{\ro{i}} \equiv (a H)_{\ro{i}}$,
and also used the relations specific to our EMDE model: $\sigma\propto 1/m^3$ and $\rho_{\phi}(a_{\rm i})\propto m^4$. 
The factor $(k / k_{\ro{i}})^3$ just represents the choice of the Fourier mode with respect to the first mode that enters the horizon during EMDE, hence we can consider it as being independent of the cosmological background.
Thus, with an earlier onset of the EMDE, the relaxation time actually becomes longer in units of the halos' dynamical time; in other words, the gravothermal process effectively becomes slower.

Finally, we remark that gravitational scatterings can also induce the relaxation of the particles in the halo.
The time scale for this would be longer than that in eq.~(\ref{eq:tr}) if, crudely speaking, the self-interaction-induced cross section is as large as \cite{Lynden-Bell:1980xip,Balberg:2002ue}
\begin{equation}
 \sigma \gtrsim 10^{-2} \frac{m}{v_s^4 \Mp^4}.
\end{equation}
Combining this with eq.~(\ref{eq:sigma}), we thus see that 
as long as the self-coupling satisfies
\begin{equation}
 \lambda \gtrsim 10^4 \left(\frac{m}{\Mp}  \right)^2,
\label{eq:xx}
\end{equation}
the relaxation time is governed by the quartic interaction. 
This condition is satisfied in most of the parameter space we explore in Section~\ref{sec:param}.

\subsection{Core evolution}\label{sec:core}

We now evaluate the mass of the core when it collapses into a black hole, using the results from Refs.~\cite{Balberg:2002ue, Koda:2011yb, Pollack:2014rja, Outmezguine:2022bhq, Gad-Nasr:2023gvf, Essig:2018pzq}.

Within a time of $\sim 10 \tau_r^{\rm NFW}$ since the halo formation, the self-interactions act to flatten the central part of the NFW profile to form a uniform-density core. 
Denoting the core radius and density respectively by $r_c$ and $\rho_c$, the mass of the core is 
\begin{align}\label{eq:Mc_def}
    M_c= \frac{4\pi}{3}\rho_cr_c^3.
\end{align}
We also define the velocity dispersion within the core as
\begin{align}\label{eq:vc_def}
    v_c^2\equiv \frac{GM_c}{r_c}.
\end{align}
We can further introduce the dynamical and relaxation time scales within the core, $\tau_{{\rm dyn},c}$ and $\tau_{r,c}$, 
in the same way as
$\tau_{\rm dyn}^{\rm NFW}$ and $\tau_{r}^{\rm NFW}$ in eqs.~\eqref{eq:tr} and \eqref{eq:tdyn}, except that the scale parameters are replaced with the core parameters.

Subsequently, the core begins to undergo gravothermal evolution, wherein an outward flow of heat is established by the ejection of particles to outer radii. 
As gravitationally bound systems have negative heat capacities,\footnote{This is understood from the virial theorem, which states that the total energy of a self-gravitating system is negative of its kinetic energy. 
Assigning a temperature according to the kinetic energy of the particles, the system has a negative heat capacity \cite{binney2008galactic}.} 
the outward heat flow leads to an increase in the velocity of particles inside the core.
Hence with the gravothermal evolution, 
$v_c$ increases while $M_c$ decreases. 
This also entails the decrease of $r_c$ and the increase of $\rho_c$, 
as one can read off from eqs.~(\ref{eq:Mc_def}) and (\ref{eq:vc_def}).

For halos that are initially optically thin (cf. eq.~(\ref{eq:thin})), 
the gravothermal evolution starts in the LMFP regime, 
i.e. $\tau_{\ro{dyn}, c} < \tau_\ro{r,c}$.
The core parameters at the onset of the LMFP gravothermal evolution are of the same order as the initial NFW scale parameters. 
In fact, Ref.~\cite{Outmezguine:2022bhq} has numerically found that they are related by\footnote{These relations depend on the detailed definitions of the onset of the LMFP regime and the edge of the core, for which we refer the reader to~\cite{Outmezguine:2022bhq}. We keep these vague in this paper as we are only interested in order-of-magnitude estimates.
}
\begin{align}\label{eq:outmez}
    r_{c}^{\rm LMFP}=0.45r_s, 
\quad
\rho_{c}^{\rm LMFP}=2.4\rho_s.
\end{align}
Here, the superscript ``LMFP'' refers to values at the onset of the LMFP evolution. 
From these relations, and also rewriting the NFW parameters in terms of the halo parameters using the results in Subsection~\ref{sec:halo}, the mass and velocity dispersion within the core can be written as
\begin{equation}
\begin{split}
   M_c^{\rm LMFP} & \sim 3\times 10^{-2}M_{\rm halo},
\label{eq:core_initial}
\\
   v_{c}^{\rm LMFP} & \sim 1 \times v_{\rm vir}.
\end{split}
\end{equation}
Similarly, the ratio between the dynamical and relaxation time scales within the core can be related to that for the initial NFW profile as
\begin{align}\label{eq:tau_lmfp}
  \frac{\tau_{\ro{dyn},c}^{\rm LMFP}}{\tau_{r,c}^{\rm LMFP}} 
 \sim 0.6 \, \frac{\tau_{\rm dyn}^{\rm NFW}}{\tau_{r}^{\rm NFW}}.
\end{align}

During most of the gravothermal evolution,
$\rho_c$ scales with $r_c$ as
\begin{align}\label{eq:scale}
    \rho_c\propto r_c^{-\alpha},
\end{align}
with a constant and positive~$\alpha$.
The scalings of the other core parameters follow as,
\begin{align}\label{eq:scale2}
 M_c\propto r_c^{3-\alpha}, && v_c^2\propto r_c^{2-\alpha}, && \frac{\tau_{\ro{dyn},c}}{\tau_{r,c}} \propto r_c^{1-\alpha}.
\end{align}
The increase of $v_c$ along with the decrease of $M_c$ during the gravothermal evolution implies that $2<\alpha<3$. 

In the LMFP regime, the evolution of the halo is self-similar with $\alpha\approx 2.190$ \cite{Lynden-Bell:1980xip,Balberg:2002ue,Outmezguine:2022bhq}. 
Consequently, the core mass shrinks rapidly as the velocity increases as $M_c\propto v_c^{-8.5}$. The mass loss continues until $v_c$ reaches $\sim 1/3$, after which relativistic instability causes the core to collapse into a black hole \cite{Balberg:2001qg,Feng:2021rst}.\footnote{This can also be seen by noting that $v_c$ of order unity corresponds to the core radius being close to the Schwarzschild radius, cf.~(\ref{eq:vc_def}).}
If the LMFP evolution continues all the way until the instability, then one finds from 
eq.~(\ref{eq:core_initial}) and $v_{\rm vir}\sim 0.5 \times 10^{-2}$ that the fraction of the halo mass collapsing into a black hole is as tiny as $\sim 10^{-16}M_{\rm halo}$.

However, since the dynamical/relaxation time ratio, $\tau_{\ro{dyn},c}/\tau_{r,c}$, increases with the gravothermal evolution, the core can enter the short mean free path (SMFP) regime, i.e. where $\tau_{\ro{dyn}, c} > \tau_{r, c}$, before collapsing into a black hole.
In the SMFP regime, the core loses its mass only by the evaporation of particles in the vicinity of the core surface, 
as opposed to the LMFP regime where particles from anywhere in the core can escape.
The mass loss from the core is hence significantly reduced during the SMFP regime, and consequently, the actual black hole mass is considerably boosted.

The transition between the LMFP and SMFP regimes takes place when 
$\tau_{\ro{ dyn}, c}/\tau_{r,c}=1$. Hence, using the last scaling relation in eq.~\eqref{eq:scale2} with $\alpha = 2.190$, the ratio between the core radii at the onset of the SMFP and LMFP evolutions is found to be
\begin{align}\label{eq:rc_smfp}
\frac{r^{\rm SMFP}_c}{r^{\rm LMFP}_c}
=
\left(
\frac{\tau_{\ro{dyn},c}^{\ro{LMFP}}}{\tau_{r,c}^{\ro{LMFP}}}
\right)^{0.84}
\sim 0.7 \left(\frac{\tau_{\rm dyn}^{\rm NFW}}{\tau_r^{\rm NFW}}\right)^{0.84}.
\end{align}
The superscript ``SMFP'' refers to quantities when the core enters the SMFP regime, and the second equality is obtained from eq.~\eqref{eq:tau_lmfp}.

The ratio of the radii in eq.~(\ref{eq:rc_smfp})
can be combined with the scalings in eq.~(\ref{eq:scale2}) to yield ratios of other core quantities during the LMFP and SMFP regimes. 
Further, using eq.~(\ref{eq:core_initial}) to connect the LMFP parameters with the initial NFW halo, one obtains the core mass and velocity dispersion at the onset of the SMFP regime as
\begin{equation}
 \begin{split}
    M_c^{\rm SMFP} & \sim 2\times 10^{-2} M_{\rm halo}\left(\frac{\tau_{\rm dyn}^{\rm NFW}}{\tau_{r}^{\rm NFW}}\right)^{0.68},
\label{eq:mass_smfp}
\\
    v_c^{\rm SMFP} & \sim 1 \times v_{\rm vir}\left(\frac{\tau_{\rm dyn}^{\rm NFW}}{\tau_{r}^{\rm NFW}}\right)^{-0.080}.
 \end{split}
\end{equation}
This explicitly shows that
cores of halos that are initially optically thinner, 
i.e. have smaller~$\tau_{\rm dyn}^{\rm NFW} / \tau_{r}^{\rm NFW}$,
loose more mass before reaching the SMFP regime. 
However, we note that the overall time until the core collapse is still roughly the same $200\tau_{r}^{\rm NFW}$ for all halos because the evolution time scale becomes shorter towards the collapse.
We also remark that, since the transition between the two regimes is set by the relation between two timescales ($\tau_{{\rm dyn},c}$ and $\tau_{r,c}$), the self-similarity present in the LMFP evolution is broken in the SMFP regime.

The scaling parameter~$\alpha$ takes different values in the LMFP and SMFP regimes. 
Numerical studies have found $\alpha$ deep in the SMFP regime to be typically close to $2.5$.
In particular, Ref.~\cite{Balberg:2002ue} claims that $\alpha$ lies between 2.52 and 2.55, whereas Ref.~\cite{Gad-Nasr:2023gvf} claims $\alpha$ to be much closer to (but still larger than) $2.5$. 
In this paper, we take $\alpha=2.5$ in the SMFP regime. Moreover, we make the approximation that $\alpha$ instantaneously transitions\footnote{Ref.~\cite{Gad-Nasr:2023gvf} tracks the gradual change in $\alpha$ to evaluate the final mass at the relativistic instability. We find that their result only differs by order 1 factors compared to our simplified calculation.} from 2.190 to 2.5 
when $\tau_{\ro{ dyn}, c} = \tau_{r,c}$.
Then, the final mass of the core at the relativistic instability can be written as
\begin{align}
    \frac{M^{\rm rel}_{c}}{M_c^{\rm SMFP}}=\left(\frac{v_{c}^{\rm SMFP}}{1/3}\right)^{2}.
\end{align}
Here the superscript ``rel'' refers to quantities at the relativistic instability when $v_c=v_c^{\rm rel}=1/3$.
Combining this with \eqref{eq:mass_smfp}, we obtain the core mass that collapses into a black hole as
\begin{align}\label{eq:seed_mass}
    M^{\rm rel}_{c} \sim 0.9 \times 10^{-5}\left(\frac{\tau_{\rm dyn}^{\rm NFW}}{\tau_{r}^{\rm NFW}}\right)^{0.52}M_{\rm halo}.
\end{align}
We remark that even if $\alpha=2.55$ is used instead of $\alpha=2.5$ in the SMFP regime, the differences in the final core mass as well as the subsequent results are insignificant.

One thus sees that the SMFP regime allows the core mass at relativistic instability to be much larger than $10^{-16}M_{\rm halo}$, which is the expectation from the LMFP evolution alone. On the other hand for $\tau_{r}^{\rm NFW} / \tau_{\rm dyn}^{\rm NFW} \gtrsim 10^{21}$, one may naively think that $M_c^{\rm rel}$ for the SMFP core becomes smaller than $10^{-16}M_{\rm halo}$. However what actually happens in such a case is that the relativistic instability kicks in before the core enters the SMFP regime, as one can directly check from~eq.~\eqref{eq:mass_smfp}\footnote{Eq.~\eqref{eq:mass_smfp} does not yield exactly $10^{21}$ as the threshold value, but this is because the coefficients have been rounded.}; 
hence the expression in eq.~(\ref{eq:seed_mass}) for
$M_c^{\rm rel}$ breaks down. 
We have checked that $\tau_{r}^{\rm NFW} / \tau_{\rm dyn}^{\rm NFW}$
is smaller than $10^{21}$ in most of the EMDE parameter space we explore in Section~\ref{sec:param}.
Hence, we ignore the possibility of reaching relativistic instability in the LMFP regime.

\section{Consequences of gravothermal collapse}\label{sec:times}

In this section, we discuss the final objects obtained after the gravothermal collapse. We first consider black hole formation at the end of gravothermal evolution and discuss their final masses. Next, we highlight how annihilations or pressure induced by the particle's self-interactions
can become important during gravothermal evolution and possibly impede black hole formation.

\subsection{Primordial black holes}

Without impediment from the particles' self-annihilations or pressure, gravothermal evolution continues as explained in the previous section, and the halo core eventually collapses into a black hole when $v_c\sim 1/3$. 
However the details of the evolution during the final moments of the collapse via the relativistic instability, as well as the gravothermal accretion that may follow, are not yet known. This prevents us from obtaining the exact value of the final black hole mass,~$M_{\rm BH}$.
In this study, we parameterize the uncertainty in the mass fraction of the halo that ends up in a black hole as
\begin{align}
    \eta\equiv \frac{M_{\mathrm{BH}}}{M_{\rm halo}}.
\end{align}
Below, we first estimate the lower and upper limits on $\eta$. Then, we provide the masses of the heaviest and the lightest black holes that can be produced during an EMDE.

\subsubsection{Efficiency of gravothermal accretion}
\label{sec:eta}

The core mass at the onset of the relativistic instability given in eq.~(\ref{eq:seed_mass}) can be rewritten in terms of the EMDE parameters by substituting eqs.~(\ref{eq:vs}), (\ref{eq:v_vir}), and (\ref{eq:rho_s}) into the NFW parameters, 
as well as using eq.~(\ref{eq:mhalo}) for the halo mass. This yields
\begin{multline}\label{eq:Mcorerel}
    M_{c}^{\ro{rel}} \sim 6 \times10^{15}\, {\rm g} \
\lambda^{1.0} 
\left(\frac{T_{\mathrm{rh}}}{\rm 100 \, MeV}\right)^{-2.5}
\left(\frac{a_{\rm rh}/a_{\rm i}}{10^{10}}\right)^{-1.2}
\\ \times\left(\frac{k/k_{\mathrm{rh}}}{10^4}\right)^{-1.4}.
\end{multline}

If only the halo core at the instability collapses into a black hole and no accretion takes place, then the final black hole mass
is simply~$M_{c}^{\ro{rel}}$.
This case gives rise to the minimum mass fraction,
\begin{multline}\label{eq:etamin}
 \eta_{\rm min}
= \frac{M_{c}^{\ro{rel}}}{M_{\ro{halo}}}
\sim 6\times10^{-7} \lambda^{1.0}
\left(\frac{a_{\rm rh}/a_{\rm i}}{10^{10}}\right)^{-1.2}
\left(\frac{k/k_{\rm rh}}{10^4}\right)^{1.6}
\\ \times
\left(\frac{T_{\rm rh}}{\rm 100~MeV}\right)^{-0.52}.
\end{multline}
Note that with increasing wave number, $\eta_{\ro{min}}$ increases while $M_{c}^{\ro{rel}}$ decreases, which is due to the scaling of the halo mass $M_{\rm halo}\propto 1/k^3$; see eq.~\eqref{eq:mhalo}.
Hence, larger halos (i.e. smaller~$k$) yield larger black holes.

The above estimate is for the initial seed mass of the black hole without taking into account its growth by accretion. However as the halo remains in virial equilibrium after the formation of the black hole, it would continue to have a negative heat capacity, and thus, the gravothermal evolution is expected to continue. Consequently, we expect the black hole mass to grow via gravothermal accretion.

A key difference between gravothermal accretion and accretion processes encountered in standard astrophysics is that no radiative processes are present in the former. Consequently, the energy generated from the accretion onto the black hole can only be carried away by particles escaping from the black hole. Thus, the black hole cannot accrete the entire halo via gravothermal accretion.

We can roughly estimate the maximum mass a black hole can accrete from energy conservation considerations. For a virialized halo, the total energy is equivalent to the kinetic energy multiplied by~$-1$. Hence, if the total energy is conserved, so is the kinetic energy. 
We thus equate the total kinetic energy of the particles making up the initial NFW halo and that of the halo at a later time when a black hole is present at the center,
\begin{align}
    K_{\rm NFW}= K_{\rm BH}+K_{\rm rem}.
\end{align}
Here, we have split the kinetic energy of the later halo into that of particles that have fallen into the black hole~$K_{\rm BH}$, and 
of the remaining particles~$K_{\rm rem}$.

We estimate $K_{\ro{BH}}$ by crudely modeling a black hole as a collection of particles with a velocity that triggers the relativistic instability, i.e. $v \sim 1/3$.
This yields
\begin{align}\label{eq:cons}
    K_{\rm BH}\sim \frac{1}{2}M_{\rm BH}
 \left( \frac{1}{3} \right)^2.
\end{align}
As $K_{\rm rem}$ is nonnegative, from the energy conservation equation, we thus find
\begin{align}
    M_{\rm BH} \lesssim 18K_{\rm NFW}.
\label{eq:18}
\end{align}
This shows that the mass of the black hole is maximized in the limit $K_{\rm rem}\rightarrow 0$; this corresponds to the case where all the outer halo particles have just barely escaped to infinity.

The total kinetic energy of the halo~$K_{\ro{NFW}}$ is related to the gravitational binding energy~$U$ through the virial theorem.
By calculating the gravitational energy of the initial NFW profile
up to the halo radius, we obtain
\begin{equation}\label{eq:K_NFW}
\begin{split}
 K_{\rm NFW}
= -\frac{1}{2}U(r_{\ro{halo}}) 
& = 2 \pi G
\int^{r_{\ro{halo}}}_0 M(r) \rho (r) r \, dr
\\
& =\frac{1}{2}F(c)M_{\ro{halo}}v_{\rm vir}^2.
\end{split}
\end{equation}
Here $M(r)$ is the mass within the radius~$r$, and 
\begin{equation}
F(c) = \frac{1}{h(c)^2} \left[
\frac{c}{2} - \frac{c}{c+1} \left\{
h(c) + \frac{2 c+ 1}{2 c+2}
\right\}
\right],
\end{equation}
which for $c=30$ gives $F\approx 2$.
Substituting eq.~(\ref{eq:K_NFW}) into eq.~(\ref{eq:18})
and using $v_{\rm vir}\sim 0.5 \times 10^{-2}$, we find the maximum value of $\eta$ to be\footnote{One can check that $\eta_{\ro{max}} > \eta_{\ro{min}}$ holds for arbitrary~$c$, given that the halo is initially optically thin.
This provides a sanity check on our computation of $\eta_{\rm max}$.
We should also remark that the computation breaks down for an extremely large~$c$ such that the expression for~$\eta_{\ro{max}}$ exceeds unity. Since there, the mean-square velocity
$\langle v^2 \rangle  = F(c) v_{\ro{vir}}^2$ 
(cf.~(\ref{eq:K_NFW}))
is also large, relativistic corrections cannot be ignored.}
\begin{align}\label{eq:opt}
    \eta_{\rm max}
\sim 9 F(c) v_{\ro{vir}}^2
\sim 10^{-3}.
\end{align}

We remark that the above derivation of $\eta_{\ro{max}}$ would break down if, for instance, the system falls out of virial equilibrium. Indeed,  
some earlier studies conjectured much higher final black hole masses through Bondi accretion \cite{Ostriker:1999ee,Hennawi:2001be,Hu:2005cd,Pollack:2014rja,Feng:2020kxv}.\footnote{Specifically, these studies focused on present-day dark matter halos with $c \sim 1$ and conjectured $\eta_{\rm max}\sim 10^{-2}$ \cite{Pollack:2014rja,Feng:2020kxv}. In contrast, 
eq.~\eqref{eq:opt} gives $\eta_{\rm max}\sim 10^{-4}$.}
In those models, particles spherically infall into the black hole under adiabatic conditions, eliminating the need for a gravothermal process to transport energy. However, such a spherical infall assumes a highly symmetric initial distribution; any deviations would require the gravothermal mechanism to remove angular momentum for the collapse to continue~\cite{1976ApJ...210..757S,Feng:2021rst}. Therefore, further study is needed to assess the initial sphericity and whether Bondi or gravothermal accretion determines the final black hole mass. In this work, we conservatively assume that the accretion is governed by the gravothermal process and adopt eq.~\eqref{eq:opt} for the maximum black hole mass.

\subsubsection{Mass limits}
\label{sec:BH_limits}

The black hole mass, $M_{BH}=\eta M_{\rm halo}$, is a decreasing function of $k$ for both $\eta=\eta_{\rm min}$ and $\eta=\eta_{\rm max}$. Thus, heavier halos always form heavier black holes. 
The largest halos that can form black holes are seen in fig.~\ref{fig:scales} as those that collapse right before reheating, 
i.e. $a_{\rm GC}=a_{\mathrm{rh}}$. 
The corresponding lower limit on $k$ has been shown earlier in eq.~\eqref{eq:kheavy}, and this sets an upper limit on $M_{\rm BH}$.
Since we are considering halos that are initially collisionless, 
we have $200\tau_{r}^{\rm NFW} \gtrsim 1/H_{\rm NL}$
(see discussions around eq.~\eqref{eq:marg}).
Thus the limit in eq.~\eqref{eq:kheavy} is set by the first term in the square brackets.
For $\eta=\eta_{\rm min}$, plugging this into eqs.~\eqref{eq:etamin} and 
\eqref{eq:mhalo} yields
\begin{align}\label{eq:Ml_pes}
    M_{\rm BH}[\eta_{\rm min}] \lesssim~&10^{16} {\rm g}\ \lambda^{1.5} \left(\frac{T_{\mathrm{rh}}}{\rm 100~MeV}\right)^{-2.8}\left( \frac{a_{\mathrm{rh}}/a_{\mathrm{i}}}{10^{10}} \right)^{-1.7}.
\end{align}
Similarly, for $\eta = \eta_{\rm max}$ (cf. eq.~\eqref{eq:opt}) we find 
\begin{align}
\label{eq:Ml}
    M_{\rm BH}[\eta_{\rm max}] \lesssim~&10^{20} {\rm g}\ \lambda \left(\frac{\eta_{\rm max}}{10^{-3}}\right)\left(\frac{T_{\mathrm{rh}}}{\rm 100~MeV}\right)^{-2.5}\nonumber\\
    &\times\left( \frac{a_{\mathrm{rh}}/a_{\mathrm{i}}}{10^{10}} \right)^{-1.1}.
\end{align}

We remind the reader that we have been setting the concentration of halos upon their formation at $a_{\rm NL}$ to the mean value $c=30$. 
If one instead keeps $c$ arbitrary, then the above upper limits on $M_{\ro{BH}}$ for $\eta = \eta_{\ro{min}}$ and $\eta_{\ro{max}}$ scale respectively as $\propto c^{2.9} $ and $c^{1.8} $, up to logarithmic corrections.\footnote{Here we also ignore the $c$-dependence of $\eta_{\ro{max}}$, since it is weak for $c \lesssim 10^2$.}
Thus, rarer halos with larger $c$ values can produce significantly heavier black holes. For brevity, in this study, we assume all halos have average properties and defer the study of variance in halos for the future.

The lightest black holes arise from the mode that enters the horizon at $a_{\mathrm{i}}$. The corresponding upper limit on the wave number is $k < (aH)_{\ro{i}}$, cf. eq.~\eqref{eq:klight}.
This translates into lower bounds on the black hole mass as,
\begin{align}\label{eq:Ms_pes}
    M_{\mathrm{BH}}[\eta_{\rm min}] & \gtrsim 10^{14} \, {\rm g}\ \lambda^{1.0}\, 
 \left(\frac{T_{\mathrm{rh}}}{\rm 100~MeV}\right)^{-2.5}\left(\frac{a_{\mathrm{rh}}/a_{\mathrm{i}}}{10^{10}} \right)^{-1.9}\!\!,
\end{align}
\begin{align}\label{eq:Ms}
    M_{\mathrm{BH}}[\eta_{\rm max}] & 
\gtrsim 10^{16} \, {\rm g}\, \left(\frac{\eta_{\rm max}}{10^{-3}}\right)\left(\frac{T_{\mathrm{rh}}}{\rm 100~MeV}\right)^{-2}\nonumber\\
    & \quad \times\left(\frac{a_{\mathrm{rh}}/a_{\mathrm{i}}}{10^{10}} \right)^{-1.5}.
\end{align}
The mass lower limit with $\eta = \eta_{\rm min}$ also depends sensitively on the concentration, as $\propto c^{2.0}$. The lower limit with $\eta = \eta_{\rm max}$, on the other hand, only has a weak $c$-dependence through~$\eta_{\ro{max}}$.

\subsection{Cannibal stars}\label{sec:cann}
So far, we have considered that the gravothermal collapse ultimately leads to the formation of a black hole by supposing the absence of any internal heat sources. 
However, 4-point interactions, which give rise to 
$2 \to 2 $ scatterings, generically also allow for $4\rightarrow 2$ annihilation reactions inside the nonrelativistic core.
(The inverse process is energetically forbidden.)
These cannibal reactions convert the rest-mass energy into kinetic energy and can provide support against gravitational collapse, similar to how nuclear fusion supports a star. 
Below we discuss the conditions for the formation of such ``cannibal stars,'' and also hypothesize on their possible collapse into black holes after accretion.

\subsubsection{Formation condition}

During the initial phase, when the halo is in the LMFP regime, we expect that cannibal annihilations have a negligible impact on the gravothermal evolution. This is because the relativistic particles produced by $4\rightarrow 2$ annihilations 
may simply escape the LMFP halo without further interactions. 
However, after entering the SMFP regime, the relativistic particles produced within the SMFP core can remain trapped, depositing their energy into the core as heat.

To compute the rate of cannibal annihilations we start by considering the Boltzmann equation. Specifically, if the distribution of the $\phi$~particles takes a Maxwell--Boltzmann form, 
the Boltzmann equation with the $ 4 \to 2$ process takes the form,
\begin{equation}
 \frac{d n_\phi}{dt} = -\left( n_\phi^4 - n_\phi^2 n_{\phi, \ro{eq}}^2 \right) 
\langle \sigma v^3 \rangle_{4 \to 2}.
\end{equation}
Here $n_\phi$ is the number density of~$\phi$, and $n_{\phi, \ro{eq}}$ is the equilibrium number density.
For the theory in eq.~\eqref{eq:phi4}, the thermally averaged annihilation cross section~$\langle \sigma v^3 \rangle_{4 \to 2}$ 
for nonrelativistic $\phi$~particles is, at tree-level,\footnote{There are discrepancies for the value of~$\langle \sigma v^3 \rangle_{4 \to 2}$ in the literature. 
For a correct derivation, see e.g.~\cite{Arcadi:2019oxh}. We note that
$\langle \sigma v^3 \rangle_{4 \to 2}$ in this paper is defined as 
that of \cite{Arcadi:2019oxh} multiplied by~$2$.}
\begin{equation}
    \langle \sigma v^3 \rangle_{4 \to 2} = \frac{1}{2048 \sqrt{3} \pi}\frac{\lambda^4}{m^8}.
\label{eq:cann}
\end{equation}
Considering that the number of particles decreases by two at each annihilation, in the following 
we use this cross-section and estimate 
the rate for the annihilation process inside the core as 
$n_\phi^4 \langle \sigma v^3 \rangle_{4 \to 2} / 2$.

To see whether the cannibal annihilations can inhibit the gravothermal evolution, one simply needs to check whether the heat produced from annihilations exceeds the heat that escapes from the collapsing SMFP core.\footnote{Annihilations also decrease the core mass. However, since the bulk of the particles within the core are nonrelativistic, the heat production affects the core evolution more significantly.}
The cannibal heating within the core per unit time is given by,
\begin{align}
    Q=\frac{\rho_{c}^4}{m^4} \frac{
 \langle \sigma v^3 \rangle_{4 \to 2}}{2}\times \frac{4\pi}{3}r_{c}^3\times 2m,
\label{eq:Q}
\end{align}
where we have replaced $n_\phi$ with $\rho_c / m$.

The heat loss in the SMFP regime can be evaluated by applying the kinetic theory of gases consisting of particles with elastic scattering interactions as
(see also \cite{Balberg:2001qg,Balberg:2002ue}),
\begin{equation}
\frac{L}{4 \pi r^2} 
\sim -\frac{v}{m\sigma} \frac{\partial T}{\partial r}
\sim -  \frac{v}{\sigma} \frac{\partial v^2}{\partial r}.
\end{equation}
Here, $L$ is the total heat flow out of a sphere at radius~$r$ per unit time, and $v$ is the velocity dispersion, which is identified with the temperature as $T \sim m v^2$.
Numerical evaluations in \cite{,Gad-Nasr:2023gvf}
found that inside an SMFP core, the velocity gradient takes a form of\footnote{From the equation of hydrostatic equilibrium,
$ \partial (\rho v^2) / \partial r \sim - G M \rho / r^2$, 
one can rewrite eq.~\eqref{eq:v2grad} as a relation between the velocity and density gradients:
$ \partial \log v^2 / \partial r \sim 
0.1 \, \partial \log \rho / \partial r$~\cite{Gad-Nasr:2023gvf}.}
\begin{equation}
\frac{\partial v^2}{\partial r }\sim - 0.1 \frac{G M(r) }{r^2},
\label{eq:v2grad}
\end{equation}
with $M(r)$ being the mass enclosed by~$r$.
Moreover, the result from \cite{Balberg:2002ue} (see their Fig.~5) suggests that the above profile can be roughly extrapolated up to the core surface, yielding the cooling rate of the SMFP core to be
\begin{equation}\label{eq:Qcool}
 L_c \equiv L(r=r_c) \sim  \frac{r_cv_c^3}{\sigma}.
\end{equation}

As long as $Q < L_c$, the heat generated from the cannibal interactions can be appropriately transferred outside of the core, allowing for the gravothermal evolution to continue. Note that the heating-to-cooling rate scales as
\begin{align}
    \frac{Q}{L_c}\propto\frac{\rho_c^4r_c^2}{v_c^3}\propto \frac{v_c^{17}}{M_c^6},
\end{align}
where in the last relation we used $\rho_c\propto M_c/r_c^3$ and $v_c^2\propto M_c/r_c$. As during gravothermal evolution, the core mass decreases while core velocity increases, the ratio~$Q/L_c$ increases.
Consequently, requiring 
\begin{equation}
\left(\frac{Q}{L_c}\right)^{\ro{rel}} \lesssim 1
\label{eq:LcQ}
\end{equation}
at the onset of the relativistic instability is sufficient to ensure that cannibal reactions remain unimportant until the formation of the black hole. 

The condition (\ref{eq:LcQ}) can be rewritten as a lower bound on the core mass. Using eqs.~(\ref{eq:Q}) and (\ref{eq:Qcool}), and further rewriting core quantities at $v_c^{\ro{rel}}= 1/3$ in terms of the core mass via eqs.~(\ref{eq:Mc_def}) and (\ref{eq:vc_def}), 
one arrives at 
\begin{align}
\label{eq:Mcann}
    M_{c}^{\rm rel}\gtrsim M_{\mathrm{can}},
\end{align}
with
\begin{equation}\label{eq:M_can}
 \begin{split}
    M_{\mathrm{can}}
& \equiv \frac{1}{2 \cdot 3^{7/3} \pi^{1/2}}
\left( \frac{\sigma \, \langle \sigma v^3 \rangle_{4 \to 2}}{G^{10} m^3} \right)^{1/6}
\\
& \approx 1 \times 10^{37} {\rm g}\ \lambda\left(\frac{\rm GeV}{m}\right)^{7/3}. 
 \end{split}
\end{equation}
Replacing $m$ in terms of the EMDE parameters (cf. eq.~\eqref{eq:mass}), we find 
\begin{multline}
    M_{\rm can}\sim 5 \times 10^{19} {\rm g}\ \lambda  \left(\frac{T_{\mathrm{rh}}}{100 \, \ro{MeV}}\right)^{-7/3}\left(\frac{a_{\mathrm{rh}}/a_{\mathrm{i}}}{10^{10}} \right)^{-7/4}.
\end{multline}
    
Note that in the above calculations, we have ignored quantum many-body effects. Specifically, at large densities, the cannibal cross-section in eq.~\eqref{eq:cann} can obtain corrections. We leave a detailed calculation of the mass threshold~$M_{\mathrm{can}}$ including such corrections for the future.

One can further convert the bound in eq.~(\ref{eq:Mcann}) into an upper limit on~$k$ by using eqs.~(\ref{eq:mass}) and (\ref{eq:Mcorerel}),
\begin{multline}\label{eq:kcan}
    \frac{k}{k_{\rm rh}} \lesssim 20 \, \lambda^{0.029} \left(\frac{T_{\mathrm{rh}}}{100 \, \ro{MeV}}\right)^{-0.13}
\left(\frac{a_{\mathrm{rh}}/a_{\mathrm{i}}}{10^{10}} \right)^{0.40}.
\end{multline}
For halos originating from wave modes that violate this bound, $Q$ becomes of order~$L_c$ before their cores collapse into black holes. 

While we have not performed detailed computations of the gravothermal process including cannibal reactions, we expect that the gravothermal evolution is halted once $L_c\sim Q$ is reached.
We could then imagine that the thermal heat from cannibal reactions provides support against gravothermal cooling and thus stabilizes the core. We call such a system a ``cannibal star''.

\subsubsection{Cannibal star accretion}

It is interesting to speculate on the possible fate of a cannibal star, in particular, whether it eventually collapses into a black hole. For this purpose, let us make a guess that the density stays uniform within the cannibal star, 
and that the relation $(\partial v^2 / \partial r)_c \sim -0.1 \, v_c^2 / r_c$ continues to hold at the edge of the star.
Then the cannibal heating and gravothermal cooling rates have the same forms as in eqs.~(\ref{eq:Q}) and (\ref{eq:Qcool}).
Further, supposing that the rates balance, i.e. $Q = L_c$, one can check that the mass of a cannibal star is related to the virial velocity within the star as $M_c \sim M_{\ro{can}} (3 v_c)^{17/6}$. 
This implies that if the surrounding halo feeds the cannibal star and allows its mass to increase continuously, then the virial velocity also increases. 
Eventually, when $v_c$ reaches $\sim 1/3$, we can expect the cannibal star to collapse into a black hole with mass~$M_{\ro{can}}$. 

The formed black hole may continue to grow with further accretion.
Hence, we can say that a black hole produced by the collapse of a cannibal star would have a final mass satisfying
\begin{equation}
 M_{\ro{BH}} \gtrsim M_{\ro{can}}.
\label{eq:73}
\end{equation}
One can think of $M_{\rm can}$ as the threshold mass needed for the cannibal star to overcome the heating and collapse into a black hole.
We also note that since a cannibal star formation entails the violation of 
the condition in eq.~(\ref{eq:Mcann}), 
this together with eq.~(\ref{eq:73}) yields $M_{\ro{BH}} > M_c^{\ro{rel}}$.
Hence a black hole arising from a cannibal star necessarily satisfies $\eta > \eta_{\ro{min}}$.

In the case of the maximum accretion, $\eta = \eta_{\rm max}$ (cf. eq.~\eqref{eq:opt}), the above mass limit gives a bound on~$k$,
\begin{multline}\label{eq:kcan2}
    \frac{k}{k_{\rm rh}} \lesssim 6\times10^3\lambda^{-0.33}
\left( \frac{\eta_{\ro{max}}}{10^{-3}} \right)^{0.33}
\left(\frac{T_{\mathrm{rh}}}{\rm 100~MeV}\right)^{0.11}\\ \times
\left(\frac{a_{\mathrm{rh}}/a_{\mathrm{i}}}{10^{10}} \right)^{0.58}.
\end{multline}
Cannibal stars arising from modes beyond this limit cannot form black holes, even if they accrete with maximum efficiency.

We emphasize that our discussion on the cannibal star evolution is based on several guesses. In reality, the core profiles may become significantly different after the formation of a cannibal star. 
Moreover, unlike accretion onto black holes, the heat from cannibal reactions may completely counter the gravothermal process and prevent any accretion.
A proper treatment would require solving a set of gravothermal equations that include cannibal heating, which is beyond the scope of this paper.

\subsubsection{Cosmological cannibalism}
\label{sec:cosmo_cann}

If cannibal interactions rapidly occur on a cosmological level,
the entire universe enters a cannibal phase \cite{Carlson:1992fn,Pappadopulo:2016pkp}, which prevents the growth of subhorizon density perturbations \cite{Erickcek:2020wzd,Erickcek:2021fsu}.
To assess whether this happens, let us use the same expressions as in the previous subsections, but with the core density~$\rho_c$ replaced by 
the cosmological background density~$\rho_\phi$.
The interaction rate per particle in the universe is then
estimated as 
$\sim (\rho_\phi / m)^3 \langle \sigma v^3 \rangle_{4 \to 2}$,
and the condition for the universe to be in an EMDE instead of a cannibal phase is
\begin{equation}
\frac{1}{H}
 \left( \frac{\rho_\phi}{m} \right)^3
\langle \sigma v^3 \rangle_{4 \to 2}
\lesssim 1.
\end{equation}
During an EMDE in which 
$\rho_\phi = 3 \Mp^2 H^2 \propto a^{-3}$, the left-hand side decreases with the cosmic expansion as 
$\propto a^{-15/2}$.
Hence, for the universe to actually be in an EMDE from $a_{\ro{i}}$~onward
as we have been supposing,
the above condition needs to hold at the energy scale~$\rho_{\phi} (a_{\ro{i}})$.
This requirement is rewritten using eq.~\eqref{eq:rho_ai} as a limit on the self-coupling,
\begin{equation}
 \lambda \lesssim 300 \left(\frac{m}{\Mp} \right)^{1/4}.
\end{equation}
Alternatively, it can be written using eq.~\eqref{eq:mass} as a limit on the reheating temperature,
\begin{equation}
T_{\rm rh} \gtrsim 1 \, \ro{GeV} \,  \lambda^4 \left(\frac{a_{\rm rh}/a_{\rm i}}{10^{10}}\right)^{-3/4}.
\label{eq:76}
\end{equation}
If the reheating temperature is below this threshold value, density perturbations inside the horizon would not start growing at~$a_{\ro{i}}$.

Before ending this subsection, we remark that number-changing annihilations can, in principle, be switched off if the particles carry a charge. 
For instance, the particles can be a complex scalar with a global U(1) symmetry, which were initially produced with an excess over the antiparticles (e.g., {\it \` a la} Affleck--Dine~\cite{Affleck:1984fy}).
Cannibalism would then be suppressed, both at the cosmological level and inside halo cores.

\subsection{Boson stars}

The gravothermal evolution studied so far does not take the quantum nature of particles into account. This can become important if the mean particle spacing inside the core, $d\sim (m/\rho_c)^{1/3}$, becomes smaller than the de Broglie wavelength of the particle, $l_{\ro{dB}}\sim 1/(mv_c)$.
The ratio of the two lengths can be written in terms of the core parameters as
\begin{align}
    \frac{l_{\rm dB}}{d}\sim \frac{\rho_c^{1/3}}{m^{4/3}v_c}=\left(\frac{3}{4\pi }\right)^{1/3}
\frac{v_c}{G m^{4/3} M_c^{2/3}},
\label{eq:75}
\end{align}
where upon moving to the far right-hand side we used 
eqs.~\eqref{eq:Mc_def} and \eqref{eq:vc_def}.
This ratio increases during gravothermal evolution since $v_c$ increases while $M_c$ decreases.
Thus, if the quantum nature is negligible at the time of the relativistic instability ($v_c=1/3$), then it would remain negligible at all previous times.
Therefore, the condition for the quantum nature of the particles to be negligible for the gravothermal collapse is
\begin{equation}
\left( \frac{l_{\ro{dB}}}{d} \right)^{\ro{rel}}
\lesssim 1.
\label{eq:dBrc}
\end{equation}
This condition can be rewritten as a lower bound on the core mass at the relativistic instability,
\begin{equation}
 M_c^{\rm rel} \gtrsim \frac{1}{6 \pi^{1/2} G^{3/2} m^2} 
\sim 3 \times 10^{32}\, \ro{g} \, \left(\frac{\rm GeV}{m}\right)^2,
\label{eq:Mg}
\end{equation}
which can further be converted into an upper limit on $k$ using eq.~(\ref{eq:Mcorerel}) as,
\begin{multline}\label{eq:k_dB}
    \frac{k}{k_{\rm rh}} \lesssim 6\times 10^2\, \lambda^{0.73}  \left(\frac{T_{\mathrm{rh}}}{\rm 100~MeV}\right)^{-0.36}   \left(\frac{a_{\mathrm{rh}}/a_{\mathrm{i}}}{10^{10}} \right)^{0.23}.
\end{multline}

If the particles are fermions, then Pauli-blocking prevents $d$ from becoming smaller than $l_{\rm dB}$. 
This implies that for halos arising from wave modes violating the bound in eq.~(\ref{eq:k_dB}), the degeneracy pressure halts further contraction. Consequently, the halo core can end up as a degenerate star. 
We do not explore the consequences of degenerate stars in this paper since our toy model for EMDE assumes bosonic particles.

\subsubsection{Quantum pressure}
For bosonic particles, a Bose condensate can form as $d$ becomes smaller than $l_{\rm dB}$. The existence of a condensate may induce some modifications to the gravothermal evolution, however, we do not expect it to prevent the core collapse itself.
For simplicity, in the following discussions, we assume that the gravothermal evolution of a condensate is the same as that in the purely classical regime.\footnote{The relaxation time is known to become shorter 
for a condensate~\cite{Levkov:2018kau}, but if it forms only towards the end of the gravothermal collapse, the overall collapse time is minimally affected. However if the scaling parameter~$\alpha$ in eq.~\eqref{eq:scale} also changes, then our discussion needs to be modified accordingly.}

For a condensed core, however, the gravothermal evolution would be halted by the wave nature of the particles once the de Broglie wavelength~$l_{\ro{dB}}$ becomes comparable to the size of the entire core, $r_c$.
At this point, the so-called quantum pressure can balance gravity~\cite{Hu:2000ke}
and lead to the formation of a stable core that is commonly known as a boson star. 
Here, note that the ratio of the length scales $ l_{\ro{dB}} / r_c \propto v_c / M_c $ increases throughout the gravothermal evolution. Hence, requiring that the wave nature of the bosonic particles does not prevent the gravothermal evolution amounts to imposing,
\begin{equation}
 \left( \frac{l_{\ro{dB}}}{r_c} \right)^{\ro{rel}} \lesssim 1.
\end{equation}
This can also be rewritten as a lower bound on the core mass \cite{Alcubierre:2003sx},
\begin{equation}
  M_c^{\rm rel} \gtrsim \frac{1}{3 G m}
\sim 0.9 \times 10^{14}\, \ro{g} \, \left(\frac{\rm GeV}{m}\right),
\label{eq:Mf}
\end{equation}
which converts into an upper limit on $k$ as
\begin{multline}
     \frac{k}{k_{\rm rh}} \lesssim 3\times 10^{10}\, \lambda^{0.73}  \left(\frac{T_{\mathrm{rh}}}{\rm 100~MeV}\right)^{-1.1}
\left(\frac{a_{\mathrm{rh}}/a_{\mathrm{i}}}{10^{10}} \right)^{-0.29}.
\label{eq:k-Mf}
\end{multline}
Note that for particles with masses smaller than the Planck scale, i.e. $ m < \Mp$, 
the condition in eq.~(\ref{eq:Mg}) is tighter than that in eq.~(\ref{eq:Mf}). 
In other words, the quantum pressure becomes relevant 
after the particle spacing is compressed below the de Broglie wavelength.

\subsubsection{Repulsive self-interaction pressure}

Besides the quantum pressure, the positive self-coupling~$\lambda$ induces a repulsive interaction, which can also provide pressure to support a boson star. 
A necessary condition for this repulsive interaction to be negligible is
that the field~$\phi_c$ describing the core condensate satisfies
\cite{Ho:1999hs}
\begin{align}\label{eq:interactions_compare}
    \frac{1}{2}m^2 \phi_c^2\gtrsim\frac{\lambda}{4!} \phi_c^4.
\end{align}
The field is related to the core density as $\rho_c \sim m^2 \phi_c^2$, hence this condition can be rewritten as \cite{Colpi:1986ye, Garani:2021gvc},\footnote{The initial EMDE density necessarily satisfies 
$\rho_{\phi} (a_{\ro{i}}) \ll 12 m^4 / \lambda$ if $\lambda \lesssim 1$, as one can check using eq.~(\ref{eq:rho_ai}).}
\begin{align}\label{eq:rhoc_boson}
  \rho_c \lesssim \frac{12m^4}{\lambda}.
\end{align}
Since the core density $\rho_c \propto v_c^6 / M_c^2$ increases during the gravothermal evolution, we impose this condition at the relativistic instability for the repulsive interaction to not halt the gravothermal evolution. 
This yields another lower bound on the core mass,
\begin{align}\label{eq:Mcom}
    M_{c}^{\rm rel}\gtrsim M_{\phi^4},
\end{align}
where\footnote{One can also derive $M_{\phi^4}$ by noting that the radius of a boson star supported by a repulsive interaction is 
$r \sim \sqrt{\lambda} \Mp / m^2$, and equating it with the Schwarzschild radius.
$M_{\phi^4}$ thus also corresponds to the critical mass when an interaction-supported boson star collapses into a black hole. 
A similar result was also obtained in \cite{Colpi:1986ye} for a complex scalar theory with a quartic potential.}
\begin{equation}\label{eq:Mphi4}
 \begin{split}
    M_{\phi^4} 
& \equiv \frac{\lambda^{1/2}}{108 \pi^{1/2} G^{3/2} m^2 }
\\
& \approx 2 \times 10^{31} \, {\rm g}\, \lambda^{1/2}\left(\frac{\rm GeV}{m}\right)^2 .
 \end{split}
\end{equation}
For $ \lambda \lesssim 1$, 
eq.~(\ref{eq:Mg}) gives a tighter limit than eq.~(\ref{eq:Mcom}). 
We also note that if $\lambda\gg (m/M_{\rm Pl})^2$, then
eq.~(\ref{eq:Mcom}) is stronger than the limit from quantum pressure in eq.~\eqref{eq:Mf}.
The expression for $M_{\phi^4}$ can be written in terms of the EMDE parameters by replacing $m$ using eq.~\eqref{eq:mass}, to yield
\begin{multline}
    M_{\phi^4}\sim 2 \times 10^{16} {\rm g}\ \lambda^{1/2} \left(\frac{T_{\mathrm{rh}}}{100 \, \ro{MeV}}\right)^{-2} \left(\frac{a_{\mathrm{rh}}/a_{\mathrm{i}}}{10^{10}} \right)^{-3/2}.
\end{multline}

The mass limit in eq.~\eqref{eq:Mcom} translates into an upper limit on $k$ as,
\begin{multline}\label{eq:kcom}
    \frac{k}{k_{\rm rh}} \lesssim 4\times10^3 \, \lambda^{0.38}  \left(\frac{T_{\mathrm{rh}}}{\rm 100~MeV}\right)^{-0.36}\left(\frac{a_{\mathrm{rh}}/a_{\mathrm{i}}}{10^{10}} \right)^{0.23}.
\end{multline}
Halos corresponding to wave numbers that violate either 
eqs.~(\ref{eq:k-Mf}) or (\ref{eq:kcom}) should form boson stars. 

We have been considering the quartic coupling~$\lambda$ to be positive, however, we should remark that a negative~$\lambda$ induces an attractive force instead of pressure. 
Most of the results in this paper can be applied to the case of a negative~$\lambda$, by making the replacement $\lambda \to  \abs{\lambda}$ in the equations. 
On the other hand, the conditions for boson star formation are modified with a negative~$\lambda$. In particular,
the attractive interaction accelerates the core collapse instead of inhibiting it \cite{Khlopov:1985fch, Guth:2014hsa}. This collapse, however, does not always lead to black hole formation---it can instead result in a ``bosenova,'' where a significant portion of the core is ejected as an outgoing stream of relativistic particles \cite{Levkov:2016rkk,Eby:2015hyx, Eby:2016cnq}.

\subsubsection{Boson star accretion}

A boson star can also eventually collapse into a black hole, given that it sufficiently accretes the surrounding $\phi$ particles. 
For boson stars supported by the repulsive interaction, the critical mass upon collapse is $M_{\phi^4}$;
hence the final black hole mass would satisfy
\begin{align}
    M_{\rm BH}\gtrsim M_{\phi^4}.
\label{eq:86}
\end{align}
Since a boson star formation entails the violation of 
the condition in eq.~(\ref{eq:Mcom}), 
a black hole arising from a boson star satisfies $\eta > \eta_{\ro{min}}$.

In the case of the maximum accretion, $\eta = \eta_{\rm max}$ (cf. eq.~\eqref{eq:opt}), the above mass bound imposes
\begin{align}\label{eq:kcom2}
    \frac{k}{k_{\rm rh}} \lesssim 8\times10^4 \, \lambda^{-0.17} 
\left(\frac{\eta_{\ro{max}}}{10^{-3}} \right)^{0.33}
\left(\frac{a_{\mathrm{rh}}/a_{\mathrm{i}}}{10^{10}} \right)^{0.50}.
\end{align}
Boson stars corresponding to wave numbers beyond this limit cannot form black holes even if they accrete with maximum efficiency.

\section{PBH abundance}\label{sec:BH}

While the halos as well as the various stars arising from their gravothermal collapse disappear as the $\phi$ particles decay, black holes can survive after the end of EMDE.
In this section, we estimate the relic PBH abundance.
Since we have been discussing the evolution of isolated halos, i.e. halos that are not significantly disrupted by external tidal forces, here we focus on the rare halos that do not merge into a bigger halo or undergo significant accretion before the gravothermal collapse.

Considering all halos that undergo a gravothermal collapse to form black holes, the black hole mass density at reheating per logarithmic mass bin is written as
\begin{equation}\label{eq:rhoBH_first}
\begin{split}
    \frac{d\rho_{\mathrm{BH}}(a_{\mathrm{rh}})}{d\log M_{\mathrm{BH}}} 
&=  M_{\rm 
 BH}\frac{dn_{\mathrm{BH}}(a_{\mathrm{rh}})}{d\log M_{\mathrm{BH}}}
\\ 
& \sim \eta\frac{d\rho_{\mathrm{halo}}(a_{\rm GC})}{d\log M_{\rm halo}}\left(\frac{a_{\rm GC}}{a_{\rm rh}}\right)^3,
\end{split}
\end{equation}
where $n_{\rm BH}$ is the physical number density of black holes up to mass~$M_{\rm BH}$.
Upon moving to the second line, we used that the number of black holes is determined by the number of halos upon their gravothermal collapse,
and subsequently the black hole number density dilutes as $d n_{\ro{BH}} \propto a^{-3}$.
We also used the approximation $d \log M_{\ro{BH}} \sim d \log M_{\ro{halo}}$, 
which holds up to order-unity factors
for the cases of $\eta = \eta_{\ro{min}}$ and $\eta = \eta_{\ro{max}}$;
this can be checked by recalling that $M_{\ro{halo}} \propto k^{-3}$, 
$\eta_{\ro{min}}\propto k^{1.6}$, $\eta_{\ro{max}}\propto k^0$,
and thus 
\begin{equation}
 \frac{d \log M_{\ro{BH}}}{d \log k}
=  \frac{d \log (\eta M_{\ro{halo}} )}{d \log k}
\sim \frac{d \log M_{\ro{halo}}}{d \log k}.
\end{equation}

For a fixed wave number~$k$, the Press--Schechter formalism tells us that at $a=a_{\rm NL}(k)$, the density of halos with mass $M_{\rm halo}(k)$ is of the same order as $\rho_{\phi}$. Afterwards, the comoving number density of halos with mass~$M_{\rm halo}(k)$ falls as $a^{-1}$ (see appendix~\ref{app:press} for details), hence
\begin{equation}
    \frac{d\rho_{\mathrm{halo}}(a)}{d\log M_{\rm halo}}\sim \rho_{\phi}(a)\frac{a_{\rm NL}}{a} 
\quad \ro{for} \, \, 
 a_{\rm NL} \leq a \leq a_{\ro{rh}}.
\end{equation}
By substituting this result into eq.~\eqref{eq:rhoBH_first} and using $\rho_\phi \propto a^{-3}$, we obtain
\begin{align}\label{eq:pbh_ab}
   \frac{d\rho_{\mathrm{BH}}(a_{\mathrm{rh}})}{d\log M_{\mathrm{BH}}} \sim \eta\rho_{\phi}(a_{\mathrm{rh}})\frac{a_{\rm NL}}{a_{\rm GC}}.
\end{align}

In the above abundance estimate, we are implicitly assuming that the halos with mass $M_{\rm halo}(k)$ at $a_{\ro{GC}}(k)$ were formed at $a_{\ro{NL}}(k)$, and hence have stayed isolated for a collapse time. In principle halos with mass $M_{\ro{halo}}(k)$ can newly form even long after $a_{\ro{NL}}(k)$, however, we expect the number of such halos to be negligibly tiny (cf. footnote~\ref{foot:26}).

\subsection{PBHs relative to dark matter}
We now compare the abundance of PBHs relative to that of dark matter. 
Supposing that the black hole density after reheating redshifts as 
$a^{-3}$ until today,\footnote{We ignore time variation of the PBH mass and number after reheating, which may be induced by mergers,
or accretion of SM particles. The effects of Hawking radiation will be discussed later.}
we find the fraction of dark matter in the present universe
that is in the form of PBHs with mass $M_{\rm BH}$ as,
\begin{equation}
\label{eq:fbh}
\begin{aligned}
    f_{\mathrm{BH}} (M_{\ro{BH}}) & \equiv  \frac{1}{\rho_{\ro{DM}}(a_0)}\left(\frac{d\rho_{\mathrm{BH}}(a_{0})}{d\log M_{\mathrm{BH}}}\right) \\
    & \sim \eta(k) \frac{\rho_{\phi}(a_{\mathrm{rh}})}{\rho_{\ro{DM}}(a_0)}\left( \frac{a_{\ro{rh}}}{a_0} \right)^3 \frac{a_{\rm NL}(k)}{a_{\rm GC}(k)},
\end{aligned}
\end{equation}
where $\rho_{\rm BH}$ is the density of black holes up to mass~$M_{\rm BH}$.
Moreover, $a_0$ denotes the scale factor today and 
$\rho_{\ro{DM}}(a_0)= 1.0 \times 10^{-11} \, \mathrm{eV}^4$
is the present-day dark matter density.

\begin{figure}
    \includegraphics[width=0.49\textwidth]{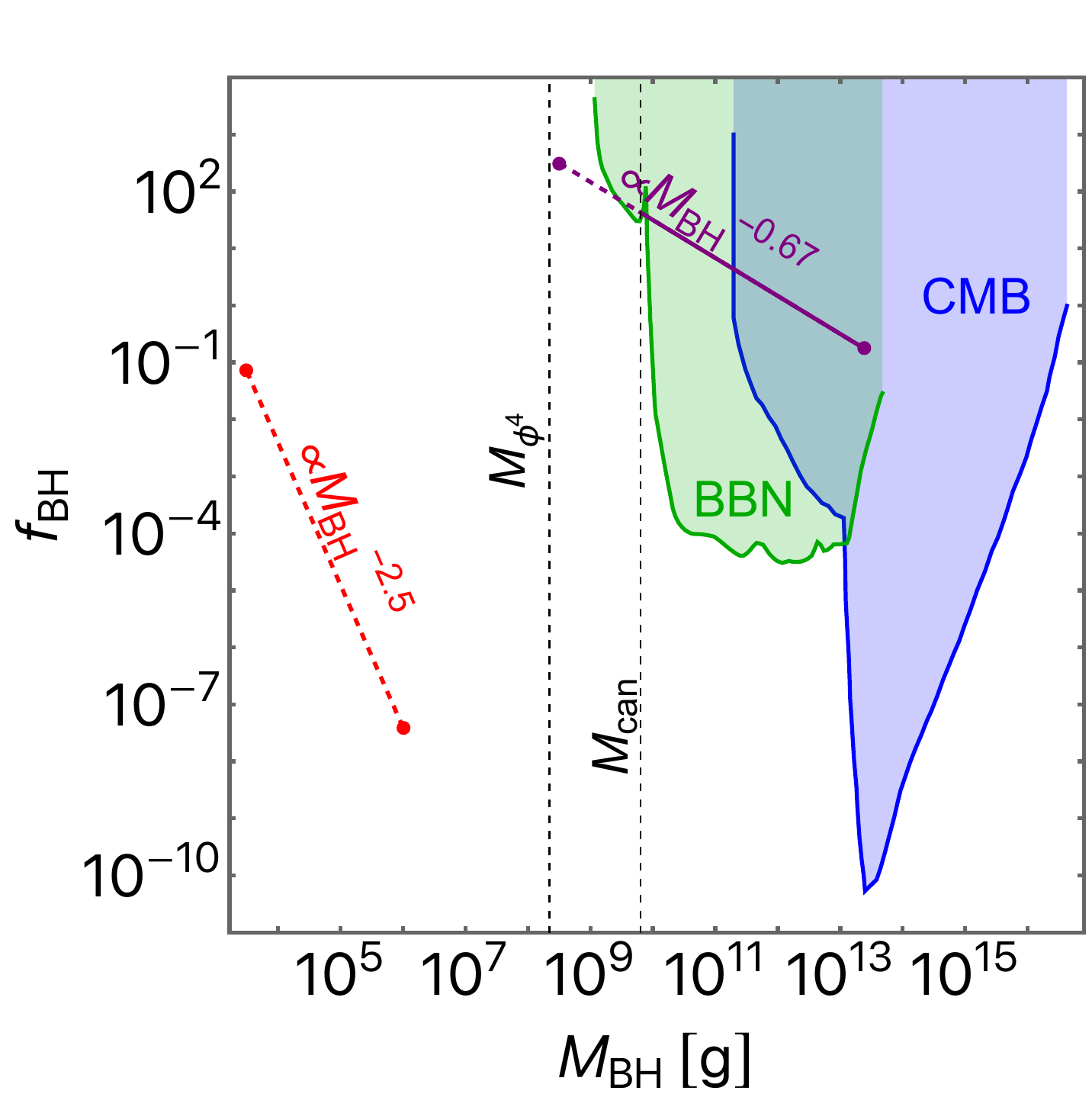}
    \caption{\label{fig:spectrum} Fraction of dark matter density in PBHs~$f_{\rm BH}$, as a function of PBH mass~$M_{\ro{BH}}$,
for a fixed parameter set of $\lambda = 10^{-1}$, $T_{\mathrm{rh}} = 100\, \mathrm{MeV}$, and $a_{\mathrm{rh}}/a_{\mathrm{i}} = 10^{15}$.
 The red line shows $f_{\rm BH}$ assuming negligible black hole accretion, while the purple line assumes maximal accretion. 
Constraints on evaporating PBHs from CMB and BBN~\cite{Carr:2009jm,Acharya:2020jbv} are shown by the blue and green regions, respectively.
The vertical dashed lines mark the threshold masses needed for the core to overcome cannibal heating or pressure from the $\phi^4$~interaction and form a black hole. See the text for details.}
\end{figure}

In the second line of eq.~(\ref{eq:fbh}), we highlighted the terms that can depend on the wave number. 
By writing $k$ in terms of $M_{\ro{BH}}$ using the results in Section~\ref{sec:eta}, we can express the PBH fraction as an explicit function of the PBH mass.
For this purpose, let us evaluate $a_{\ro{GC}}$ using the first term in the square brackets in eq.~(\ref{eq:abh}) 
by considering PBHs formed in halos that were initially collisionless
(cf. discussions around eq.~\eqref{eq:marg}).
Moreover, we calculate the redshift at reheating by assuming entropy conservation from reheating onward as,
\begin{equation}
 \frac{a_0}{a_{\ro{rh}}} \sim 1 \times 10^{12} \left( \frac{T_{\ro{rh}}}{100\, \ro{MeV}} \right).
\end{equation}
Then for the case without accretion, i.e. $\eta=\eta_{\rm min}$ (eq.~\eqref{eq:etamin}), we find
\begin{equation}
\label{eq:fbh_pes}
\begin{aligned}
    f_{\mathrm{BH}}[\eta_{\rm min}]  \sim~& \lambda^{5.0}\left( \frac{a_{\mathrm{rh}}/a_{\mathrm{i}}}{10^{10}} \right)^{-5.6} 
     \left( \frac{T_{\mathrm{rh}}}{100~\mathrm{MeV}} \right)^{-6.4} \\
    & \times\left( \frac{M_{\mathrm{BH}}}{ 10^{16}~\mathrm{g}} \right)^{-2.5}
    .
\end{aligned}
\end{equation}
For the case with the maximum accretion, i.e. $\eta=\eta_{\rm max}$ (eq.~\eqref{eq:opt}), we find
\begin{equation}
\label{eq:fbh_opt}
\begin{aligned}
    f_{\mathrm{BH}}[\eta_{\rm max}] \sim~& 10^6~\lambda^{1.3}
    \left(\frac{\eta_{\rm max}}{10^{-3}}\right)^{1.7}\left( \frac{a_{\mathrm{rh}}/a_{\mathrm{i}}}{10^{10}} \right)^{-1.5} \\
    & \times \left( \frac{T_{\mathrm{rh}}}{100~\mathrm{MeV}} \right)^{-1} \left( \frac{M_{\mathrm{BH}}}{ 10^{16}~\mathrm{g}} \right)^{-0.67}
    .
\end{aligned}
\end{equation}

If we keep the concentration~$c$ arbitrary, the abundance scales as $f_{\ro{BH}} \propto c^{9.4}$ and $\propto c^{2.3}$ for $\eta = \eta_{\rm min}$ and $\eta_{\rm max}$, respectively. These sharp $c$~dependences suggest that rarer halos with larger $c$ can have a disproportionately larger impact on the final abundance of PBHs. To accurately incorporate the contribution from rarer halos, one would need to model the distribution of halos as a function of $c$, which we leave to future work.

To give an idea of the PBH mass distribution, 
in fig.~\ref{fig:spectrum} we show the spectrum of $f_{\mathrm{BH}}$ for a fixed set of parameters: 
$\lambda = 10^{-1}$, $T_{\mathrm{rh}} = 100\, \mathrm{MeV}$, and $a_{\mathrm{rh}}/a_{\mathrm{i}} = 10^{15}$.
The red line shows $f_{\mathrm{BH}}$ as a function of $M_{\ro{BH}}$ for $\eta = \eta_{\ro{min}}$, and the purple line for $\eta = \eta_{\ro{max}}$. The endpoints of each line denote the upper and lower limits of the PBH mass, given respectively in 
eqs.~\eqref{eq:Ml_pes} and \eqref{eq:Ms_pes} for $\eta = \eta_{\ro{min}}$,
while eqs.~\eqref{eq:Ml} and \eqref{eq:Ms} for $\eta = \eta_{\ro{max}}$.
One immediately sees that the PBH spectrum is highly sensitive to the accretion scenario.
We note that the allowed mass range as well as the amplitude of $f_{\rm BH}$ can change considerably for different choices of the EMDE parameters. 
We explore the parameter space in the next section.

The vertical dashed lines mark the mass thresholds $M_{\rm can}$ for cannibal stars (cf. eq.~\eqref{eq:M_can}), and $M_{\phi^4}$ for boson stars (cf. eq.~\eqref{eq:Mphi4}).
With the chosen set of parameters in the plot, the entire range for $M_{\ro{BH}} [\eta_{\ro{min}}]$ falls short of both star thresholds.
This means that, since $M_{\ro{BH}} [\eta_{\ro{min}}]$ is equivalent to the core mass upon relativistic instability~$M_c^{\ro{rel}}$, 
the conditions in eqs.~(\ref{eq:Mcann}) and (\ref{eq:Mcom}) are violated; hence, all halos first form either cannibal or boson stars.
These stars can subsequently collapse into black holes if they sufficiently accrete to overcome the mass thresholds, cf. eqs.~(\ref{eq:73}) and (\ref{eq:86}).
Considering all stars above the thresholds to collapse, the relic abundance of PBHs thus formed is also given by $f_{\ro{BH}}$ derived above. 
In the plot, the solid part of the purple line corresponds to the spectrum of PBHs formed from the collapse of cannibal/boson stars that have accreted with maximum efficiency.
However, if the cannibal interaction or the pressure from the $\phi^4$~interaction are somehow switched off, then the PBH spectrum extends to smaller masses along the dashed lines.

We have thus far ignored Hawking evaporation, which actually gives the black holes a finite lifetime roughly of (see \cite{Page:1976df,MacGibbon:1991tj} for detailed analyses):
\begin{align}
\label{eq:tev}
    \tau_{\mathrm{ev}} \sim \, 1\, {\rm sec} \left(\frac{M_{\mathrm{BH}}}{10^9\, {\rm g}}\right)^3.
\end{align} 
In particular, PBHs with mass $M_{\ro{BH}} \lesssim 10^{15}\, \ro{g}$ evaporate by today. For such PBHs, $f_{\rm BH}$ can be considered as 
the ratio between the PBH and dark matter densities before the evaporation,
since the density ratio stays constant as long as both densities redshift as~$a^{-3}$.

In fig.~\ref{fig:spectrum}, all PBHs along the purple line evaporate by today. Hence, $f_{\ro{BH}}$ exceeding unity on the line does not mean that the PBHs overdominate the universe.
However, the radiation released from PBHs with masses within the range $10^9 \, \ro{g} \lesssim M_{\ro{BH}} \lesssim 10^{15} \, \ro{g}$  can spoil the successful predictions of BBN, or alter the anisotropies in the cosmic microwave background (CMB).
The constraints from CMB and BBN on evaporating PBHs~\cite{Carr:2009jm,Acharya:2020jbv} are shown by the blue and green regions in fig.~\ref{fig:spectrum}.

\subsection{Evaporating PBHs relative to radiation}

The constraints from CMB and BBN do not apply to PBHs that evaporate before BBN, i.e. with masses $M_{\mathrm{BH}} \lesssim 10^9\, \ro{g}$. For such PBHs, it is more informative to compare their abundance with the energy density of the SM particles, $\rho_{\ro{SM}}$, rather than the dark matter density, which is negligible before BBN. 
Here, during radiation domination, the density of PBHs within a mass bin grows relative to the SM energy density roughly as $d \rho_{\rm BH}/\rho_{\ro{SM}}\propto a$.

Hence, for PBHs that evaporate after reheating but before the matter-radiation equality, we look at the PBH density fraction at the time of the evaporation,
\begin{equation}\label{eq:betadef}
 \begin{split}
  \kappa_{\rm BH} (M_{\ro{BH}}) 
& \equiv \frac{1}{\rho_{\mathrm{SM}}(a_{\mathrm{ev}})}\left(\frac{d\rho_{\mathrm{BH}}(a_{\mathrm{ev}})}{d\log M_{\mathrm{BH}}}\right)
\\
& \sim \eta(k)  \frac{a_{\mathrm{ev}}}{a_{\mathrm{rh}}}
\frac{a_{\rm NL}(k)}{a_{\rm GC}(k)}.
 \end{split}
\end{equation}
Here $a_{\mathrm{ev}}$ is the scale factor when 
PBHs with mass~$M_{\mathrm{BH}}$ evaporate.
Upon moving to the second line, we used eq.~\eqref{eq:pbh_ab} with
$\rho_{\phi}(a_{\mathrm{rh}}) \approx \rho_{\mathrm{SM}}(a_{\mathrm{rh}})$, 
and also the scalings $d \rho_{\rm BH}\propto a^{-3}$ and $\rho_{\rm SM}\propto a^{-4}$. (We ignore time variations of $g_{*(s)}$ between reheating and PBH evaporation.)

As here we are focusing on PBHs that evaporate deep in the radiation-dominated era, the Hubble rate at the evaporation is related to the PBH lifetime via $H_{\ro{ev}} \approx 1 / 2 \tau_{\ro{ev}}$.
This Hubble rate can be rewritten in terms of the cosmic temperature upon evaporation, which in turn is related to the reheat temperature through $T_{\ro{ev}} / T_{\ro{rh}} = a_{\ro{rh}} / a_{\ro{ev}}$.
Hence, the redshift ratio can be expressed in terms of the PBH mass as,
\begin{align}\label{eq:aev}
    \frac{a_{\mathrm{ev}}}{a_{\mathrm{rh}}} \sim 2\times 10^2 
\left( \frac{T_{\mathrm{rh}}}{100\, \mathrm{MeV}} \right) 
\left( \frac{M_{\mathrm{BH}}}{10^9\, \mathrm{g}} \right)^{3/2} .
\end{align}
Using this and following the same steps as we did for $f_{\ro{BH}}$, 
one can write $\kappa_{\ro{BH}}$ as a function of $M_{\ro{BH}}$ for the cases of $\eta = \eta_{\ro{min}}$ and $\eta_{\ro{max}}$ respectively as,
\begin{equation}
\label{eq:betapes}
\begin{aligned}
    \kappa_{\rm BH}[\eta_{\rm min}] \sim & 10^{-16} \, \lambda^{5.0} \left( \frac{a_{\mathrm{rh}}/a_{\mathrm{i}}}{10^{15}} \right)^{-5.6} \\
    & \times 
\left( \frac{T_{\mathrm{rh}}}{100~\mathrm{MeV}} \right)^{-6.4}
  \left( \frac{M_{\mathrm{BH}}}{10^{9}~\mathrm{g}} \right)^{-0.98} ,
\end{aligned}
\end{equation}
\begin{equation}
\label{eq:betaopt}
\begin{aligned}
    \kappa_{\rm BH}[\eta_{\rm max}] \sim & 10^{-3} \, \lambda^{1.3}  \left(\frac{\eta_{\rm max}}{10^{-3}}\right)^{1.7}
\left( \frac{a_{\mathrm{rh}}/a_{\mathrm{i}}}{10^{15}} \right)^{-1.5}
 \\
    & \times \left( \frac{T_{\mathrm{rh}}}{100~\mathrm{MeV}} \right)^{-1} \left( \frac{M_{\mathrm{BH}}}{10^{9}~\mathrm{g}} \right)^{0.83} .
\end{aligned}
\end{equation}

Note that for $\eta = \eta_{\ro{max}}$, the density ratio~$\kappa_{\ro{BH}}$ between PBHs and SM particles increases with $M_{\rm BH}$.
This is in contrast to $f_{\ro{BH}}$, which corresponds to the fraction of dark matter in PBHs upon evaporation (cf. discussion below eq.~(\ref{eq:tev})), being a decreasing function of $M_{\ro{BH}}$.
This inversion of the mass dependence is a consequence of heavier black holes evaporating later and the density of radiation redshifting faster than that of dark matter.
The result shows that the later evaporation of heavier black holes more than compensates for their lower number densities. 
This, however, is not the case in the absence of accretion.
Here $\eta_{\ro{min}}$ sharply decreases with $M_{\rm BH}$, and consequently $\kappa_{\ro{BH}}$ also is a decreasing function of~$M_{\rm BH}$.

If the PBHs come to dominate the universe before evaporating, i.e. $\kappa_{\rm BH}>1$, our estimates of $f_{\rm BH}$ in eqs.~\eqref{eq:fbh_pes} and (\ref{eq:fbh_opt}) break down, as the assumption of entropy conservation after EMDE no longer holds.
For the parameter set in fig.~\ref{fig:spectrum}, one can check that 
$\kappa_{\rm BH}$ is smaller than unity in the entire allowed PBH mass range.\footnote{$\kappa_{\ro{BH}}$ is related to $f_{\ro{BH}}$ via 
\begin{equation*}
 \kappa_{\rm BH}
 = \frac{\rho_{\rm DM}(a_0)[a_0/a_{\rm ev}]^3}{\rho_{\mathrm{SM}}(a_{\mathrm{ev}})}f_{\rm BH}
\sim 1 \times 10^{-6} \left( \frac{M_{\mathrm{BH}}}{10^9\, \mathrm{g}} \right)^{3/2}f_{\rm BH}.
\label{eq:102}
\end{equation*}}
For PBHs that evaporate between BBN and matter-radiation equality, $\kappa_{\rm BH}$ is constrained to be much smaller than one by the BBN and CMB measurements. 

Before ending this section, we should 
remark that our abundance estimate ignores the complex mass assembly history of halos. Halos actually grow through the continued merger and accretion of smaller objects, which can affect their gravothermal evolution. Likewise, the accretion of the formed black holes can be affected by mergers of their host halos.\footnote{Our derivation of the maximum mass parameter~$\eta_{\ro{max}}$ is based on the assumption that the host halo stays isolated. Hence, in principle, the mass parameter of a black hole formed in a halo that subsequently undergoes a major merger may exceed~$\eta_{\ro{max}}$.}
Thus, halos arising from the same perturbation mode can produce black holes with different masses, depending on their mass assembly histories.
In this work, for simplicity, we ignore such dispersions.

We also ignored black hole formation in subhalos, i.e. halos that are accreted into larger ones before $a_{\rm GC}$. 
However, we expect a gravothermal collapse to also occur in rare subhalos that formed much earlier than their host halos, 
since such subhalos would be much denser than the host and tidal effects might be negligible.
This may significantly enhance the PBH abundance.
However, detailed numerical studies will be necessary to analyze the gravothermal evolution of subhalos.\footnote{Such studies have been preformed for self-interacting dark matter \cite{Nishikawa:2019lsc, Sameie:2019zfo, Kahlhoefer:2019oyt, Correa:2020qam, Kummer:2017bhr, Nadler:2020ulu, Slone:2021nqd, Zeng:2021ldo}, and the recent work \cite{Zeng:2021ldo} found that tidal heating completely suppresses the evolution. However, these studies focus on present-day subhalos, whose formation redshift is separated from that of the host by at most $\sim 10$.
A much larger separation can be realized during EMDE, which may suppress the impact of tidal heating.}

\section{Parameter space}\label{sec:param}
In this section, we outline the EMDE parameter space, i.e. $\{T_{\rm rh},\ a_{\rm rh}/a_{\rm i},\ \lambda\}$, where the gravothermal phenomena produce various compact objects. 

\begin{figure*}[pht!] 
\begin{subfigure}{0.45\textwidth}
    \includegraphics[width=1.00\textwidth]{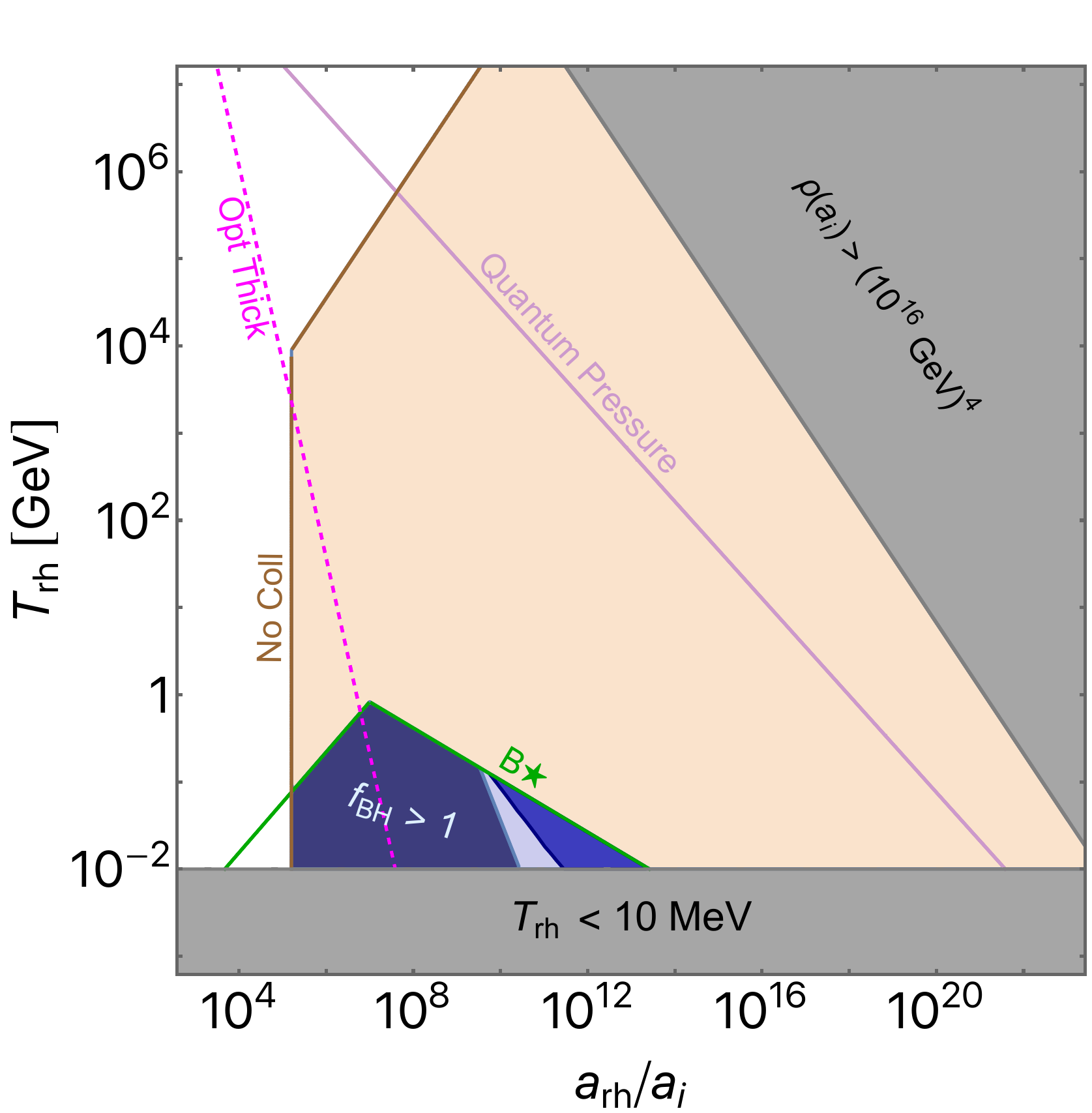}
    \caption{$\lambda = 1,\ \eta=\eta_{\rm min}$, without cannibalism.}
    \label{fig:PBH_3pes}
\end{subfigure}
\begin{subfigure}{0.45\textwidth}
    \includegraphics[width=1.00\textwidth]{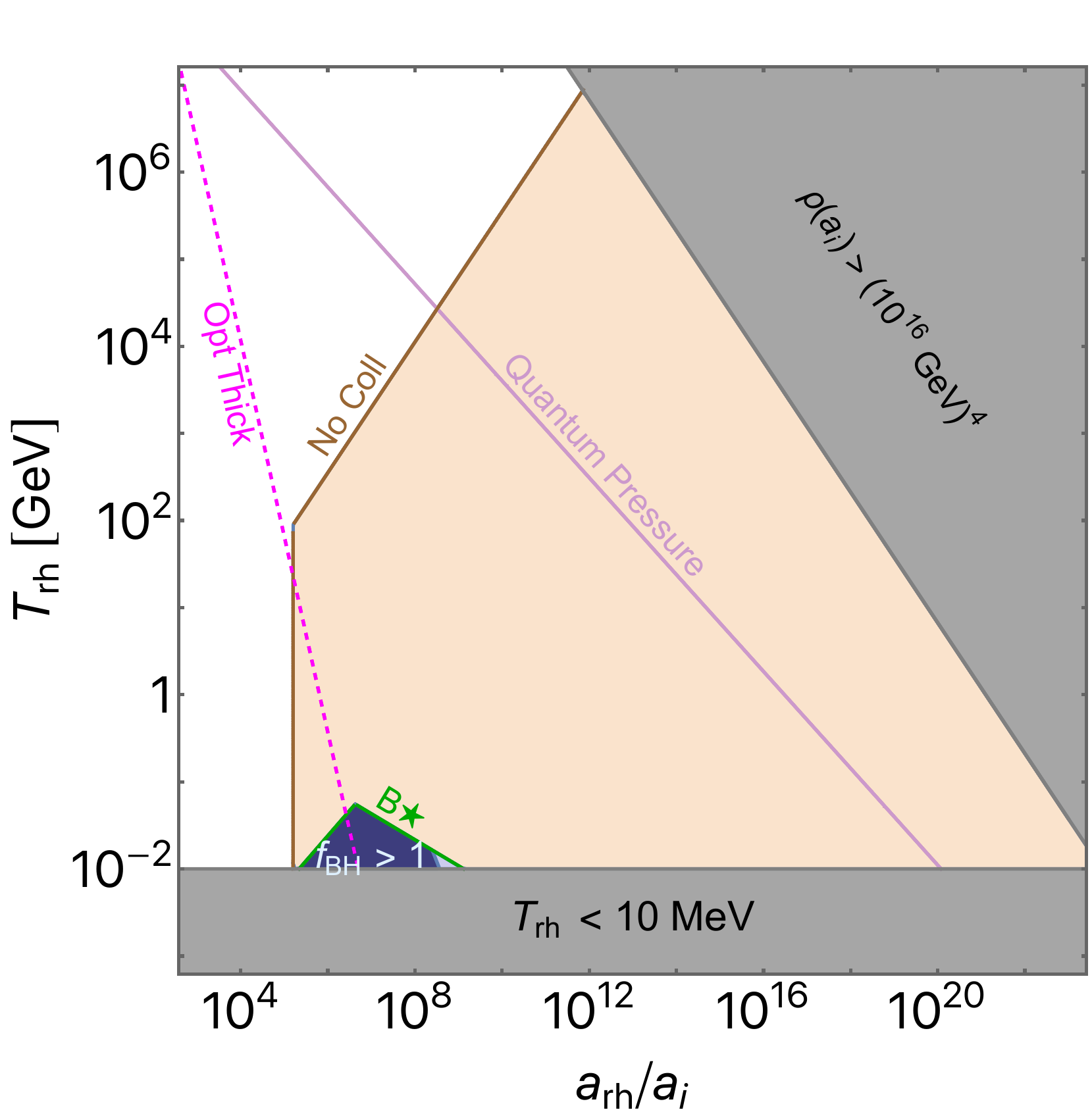}
    \caption{$\lambda = 10^{-1},\ \eta=\eta_{\rm min}$, without cannibalism.}
    \label{fig:PBH_5pes}
\end{subfigure}

\begin{subfigure}{0.45\textwidth}
\includegraphics[width=1.00\textwidth]{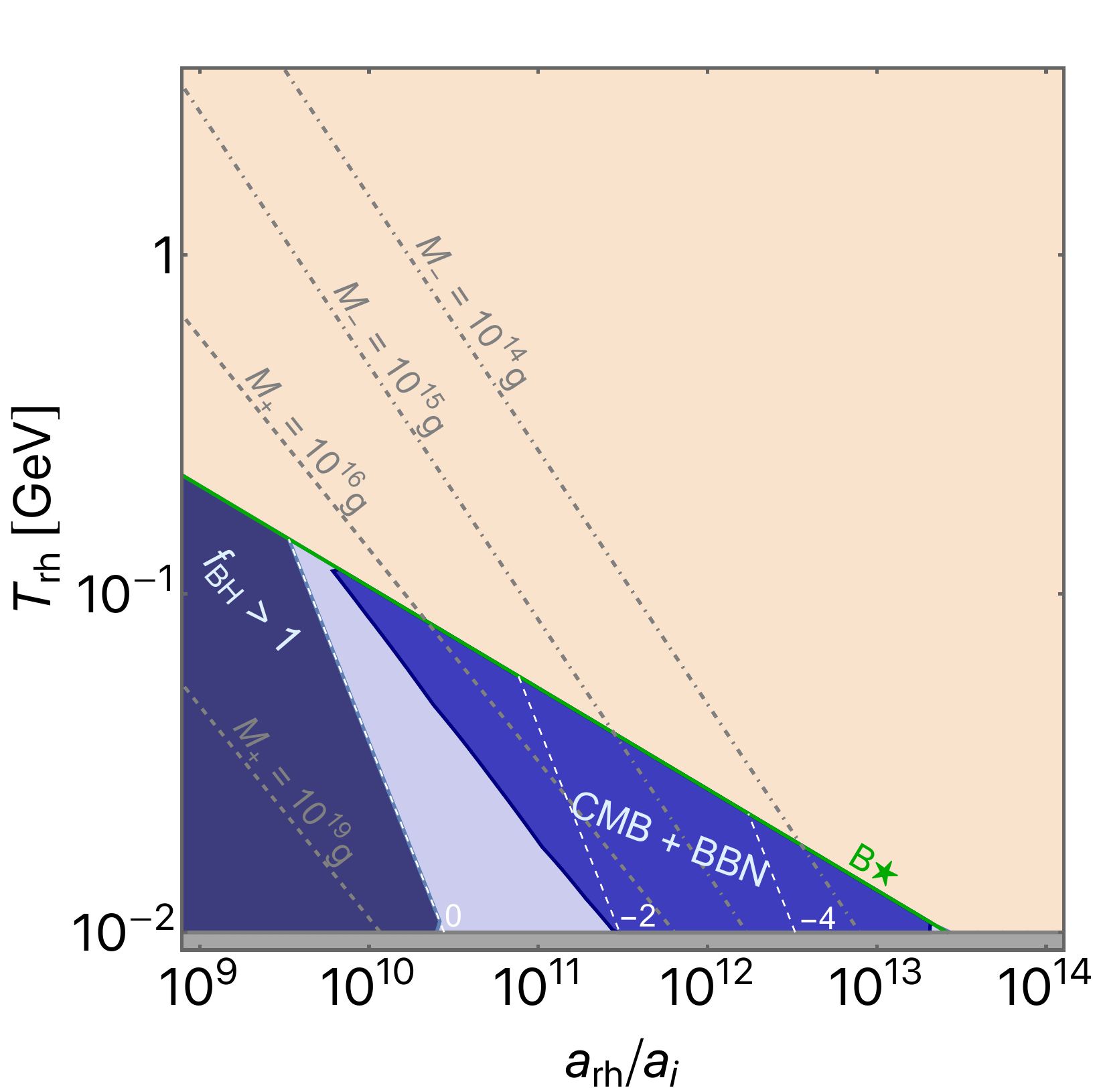}
   \caption{Zoom-in of (a).}
    \label{fig:PBH_3pes_zoom}
\end{subfigure}
\begin{subfigure}{0.44\textwidth}
    \includegraphics[width=1.00\textwidth]{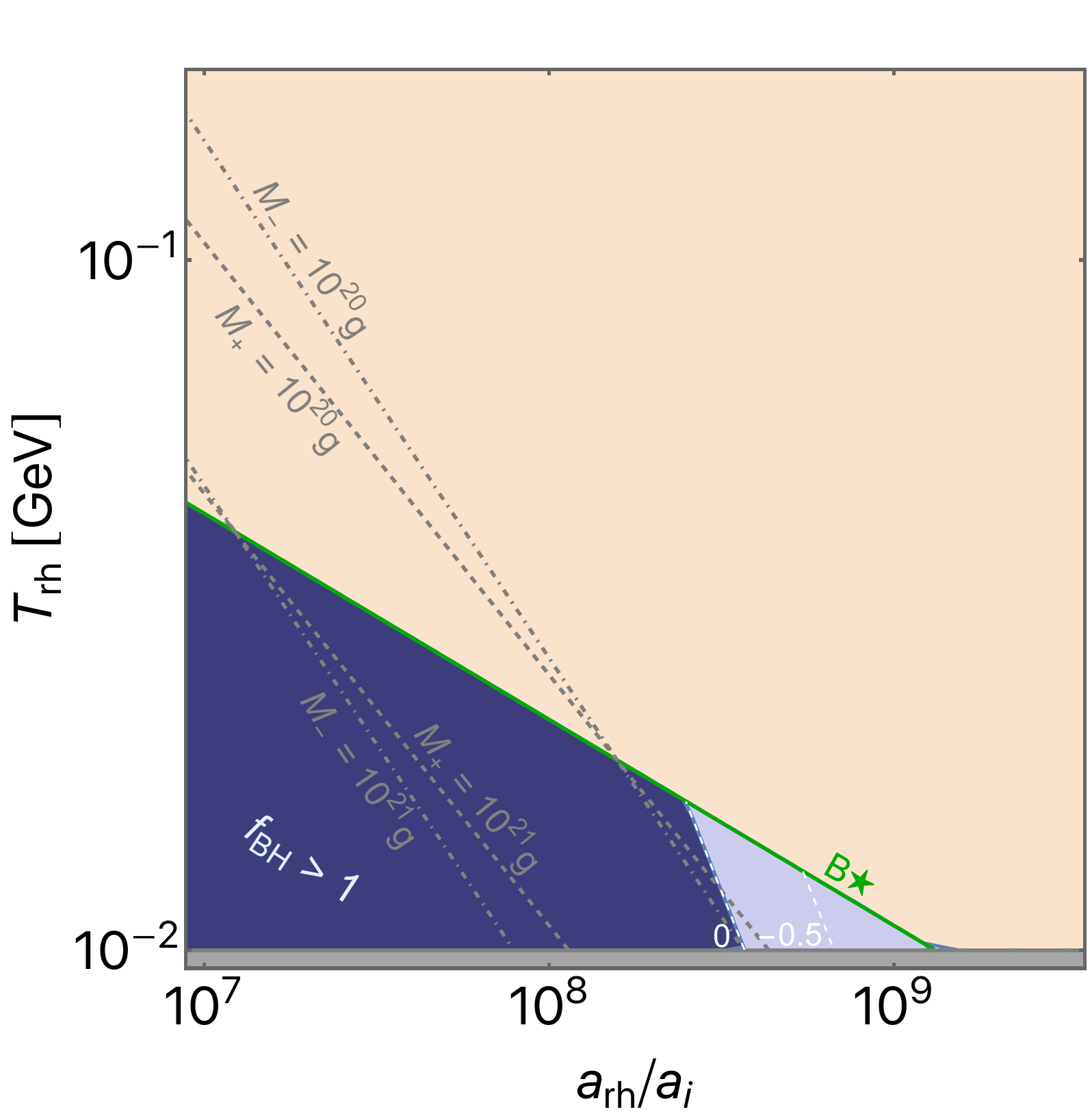}
   \caption{Zoom-in of (b).}
    \label{fig:PBH_5pes_zoom}
\end{subfigure}
\caption{Consequences of gravothermal catastrophe in terms of reheat temperature~$T_{\rm rh}$, and duration of EMDE~$a_{\rm rh}/a_i$, with  
minimal gravothermal accretion.
The self-coupling $\lambda$ is varied in the left and right panels, while cannibal annihilations are assumed to be negligible.
\textbf{Top Panels}:
The gray regions are excluded by requiring EMDE to be after inflation and before BBN. Gravothermal collapse does not occur 
on the left side of the line labeled ``No Coll.''
All collapsing halos on the left of ``Opt Thick'' are initially optically thick.
In the orange region, all halos collapse into boson stars; these are supported by the repulsive self-interaction above ``B$\star$,'' and also by the quantum pressure above ``Quantum Pressure.'' 
PBHs form in the region shown in various shades of blue. 
\textbf{Bottom Panels}: 
Zoom-in of the region where PBHs form. 
The white dashed contours show the PBH fraction in dark matter,
$\log_{10} f_{\ro{BH}}$,
while the gray dot-dashed and dashed contours respectively show the minimum ($M_-$) and maximum ($M_+$) PBH masses.
The PBHs overdominate the present universe in the region labeled ``$f_{\ro{BH}} > 1$,'' 
while the region with ``CMB+BBN'' is ruled out by CMB and BBN constraints on evaporating PBHs. 
The light blue region produces PBHs that are not in conflict with observations. See the text for details.}
\label{fig:nocan_abundance}
\end{figure*}

We start by listing some basic requirements for an EMDE. 
Firstly, since it takes place after the end of inflation,
the energy scale at the onset of EMDE is bounded from above by the CMB upper limit on the inflation scale~\cite{Planck:2018jri} as
$ \rho_{\phi}(a_{\mathrm{i}})<(1.6\times 10^{16} \, {\rm GeV})^4$.
This can be rewritten using eqs.~\eqref{eq:rho_ai} and \eqref{eq:mass} as
an upper limit on the reheating temperature,
\begin{align}\label{eq:inflation2}
    T_{\rm rh} \lesssim 2\times 10^{8} \, \ro{GeV} 
\left(\frac{a_{\rm rh}/a_{\rm i}}{10^{10}}\right)^{-3/4}.
\end{align}
An EMDE should also end before BBN. In particular, requiring that it does not alter the abundance of light elements produced during BBN or  the anisotropies in the CMB imposes \cite{deSalas:2015glj,Hasegawa:2019jsa}
\begin{align}\label{eq:BBN}
    T_{\mathrm{rh}} \gtrsim 10~{\rm MeV}.
\end{align}
From these two limits on $T_{\ro{rh}}$, one also sees that the duration of EMDE is bounded as
\begin{equation}
 \frac{a_{\rm rh}}{a_{\rm i}} \lesssim 6 \times 10^{23}.
\label{eq:106}
\end{equation}
Additionally, for the universe to enter 
an EMDE instead of a cannibal phase at~$a_{\ro{i}}$, 
the reheat temperature also needs to satisfy the lower limit in eq.~(\ref{eq:76}).

As discussed in the previous sections, certain conditions are required for an EMDE to host halos that undergo a gravothermal collapse.\footnote{For the values of~$\lambda$ chosen in the plots below, the condition for negligible gravitational scatterings in eq.~(\ref{eq:xx}) 
is satisfied in the allowed parameter window, except in regions where eq.~(\ref{eq:inflation2}) is nearly saturated. We thus ignore gravitational scatterings.} 
In the EMDE model under consideration, perturbation modes with
$ k < (a H)_{\ro{i}}$ (cf. eq.~\eqref{eq:klight})
can form halos, which undergo a gravothermal collapse before reheating
if $k$ also satisfies the lower limit in eq.~\eqref{eq:kheavy}. 
Hence, for gravothermal collapse to occur at all, these limits on the two ends of the wave number should allow for a window to exist.
This requires that the EMDE parameters satisfy both 
\begin{align}\label{eq:No_coll1}
    T_{\rm rh} \lesssim 4\times 10^7 \, \ro{GeV} \,  \lambda^2\left(\frac{a_{\rm rh}/a_{\rm i}}{10^{10}}\right)^{3/4},
\end{align}
and
\begin{align}\label{eq:No_coll2}
    a_{\rm rh} \gtrsim 1 \times 10^5 a_{\mathrm{i}},
\end{align}
where the first (second) condition arises from the first (second) term in the square brackets in eq.~\eqref{eq:kheavy}.

We note that among the $k$-dependent conditions we have introduced,
only eq.~\eqref{eq:kheavy} gives a lower limit.\footnote{We have been assuming that a halo core does not reach relativistic instability in the LMFP regime, i.e., 
$\tau_{r}^{\rm NFW} / \tau_{\rm dyn}^{\rm NFW} \lesssim 10^{21}$
(see discussions below eq.~(\ref{eq:seed_mass})), which also translates into a lower limit on~$k$.
This however is weaker than eq.~\eqref{eq:kheavy} 
in the allowed parameter windows in the plots below, 
except for in a very small region at the bottom 
right corner of fig.~\ref{fig:PBH_abundance4opt}.}
We will hence often compare it with various upper limits on~$k$ 
in the discussions below.

For $\lambda \gtrsim 10^{-10}$,
there exists an EMDE parameter window where all of 
eqs.~(\ref{eq:inflation2})--(\ref{eq:No_coll2}) are satisfied.
Here, gravothermal collapse can occur without conflicting with BBN or inflation constraints and give rise to a variety of compact objects, including PBHs, boson stars, and cannibal stars.
We focus primarily on PBHs, as they hold promise to produce observable signals.
To capture the range of possible outcomes, we first examine the parameter space assuming no accretion and then with maximal accretion.

\subsection{Minimal gravothermal accretion}\label{sec:param_possible}

We start by assuming negligible accretion, i.e. $\eta=\eta_{\rm min}$ (see eq.~\eqref{eq:etamin}). 
In this case,
boson and cannibal stars do not surpass their mass thresholds for collapsing into PBHs. 
Hence, PBHs can only be produced from the direct gravothermal collapse of the initial halo.

Fig.~\ref{fig:nocan_abundance} shows the parameter space in the plane of 
$a_{\rm rh}/a_{\mathrm{i}}$ and $T_{\rm rh}$, 
with $\lambda$ fixed to 1 (left panels) and $10^{-1}$ (right). 
The gray-shaded regions are disallowed by the BBN and inflation constraints, i.e. eqs.~(\ref{eq:inflation2}) and (\ref{eq:BBN}).
Gravothermal collapse occurs in the region to the right of the line labeled ``No Coll'', where the conditions in eqs.~\eqref{eq:No_coll1} and \eqref{eq:No_coll2} are satisfied. 
On the left of the line labeled ``Opt Thick'', 
all collapsing halos are initially optically thick;
there the upper limit on~$k$ in eq.~\eqref{eq:knfw} for halos to be initially collisionless, is incompatible with
the lower limit in eq.~\eqref{eq:kheavy} for collapsing before reheating.

Within the entire parameter region where a collapse takes place, 
we find that cannibal annihilations cannot be ignored.
This can be seen by comparing the condition for negligible cannibal annihilations in eq.~\eqref{eq:kcan} 
with eq.~\eqref{eq:kheavy}:
The necessary condition for the two to be compatible can be obtained by using just the first term in eq.~\eqref{eq:kheavy} as,
\begin{align}\label{eq:boundary_can}
    T_{\rm rh} \lesssim 2 \times 10^{-9} \, \ro{GeV} \, 
\lambda^{1.2}\left(\frac{a_{\rm rh}/a_{\rm i}}{10^{10}}\right)^{0.091}.
\end{align}
This, combined with $\lambda \ll 16\pi^2/3$ for loop corrections to the quartic coupling to be negligible, 
and the bound on $a_{\rm rh}/a_{\rm i}$ in eq.~(\ref{eq:106}), 
requires an unrealistic reheating temperature below the BBN scale. 
Thus, for the toy model under consideration, PBH formation is generically inhibited by cannibal annihilations.
However, the annihilations may be switched off by 
considering models where particle number is conserved (see the discussion below eq.~(\ref{eq:76})).
Hence, hereafter, we assume that cannibalism is suppressed.

We also find that, within almost the entire parameter windows displayed, the condition in eq.~(\ref{eq:k_dB}) 
for the particle spacing to be smaller than the de Broglie wavelength,
is incompatible with eq.~\eqref{eq:kheavy}.
Hence, most collapsing halos form Bose condensates.
A condensed core can still form a PBH, however, 
in the region above the line labeled~``B$\star$'',
all collapsing cores end up in boson stars; here the condition in eq.~\eqref{eq:kcom} for the repulsive $\phi^4$ interaction to be negligible is incompatible with eq.~\eqref{eq:kheavy}.\footnote{Recall that the two limits on~$k$ in eq.~\eqref{eq:kheavy} are, roughly, for halos that are initially optically thin or thick.
This is why the ``B$\star$'' and some other lines bend at around where they cross the ``Opt Thick'' line.
As we do not have a recipe for computing the evolution of initially thick halos, the bounds shown on the left of ``Opt Thick'' should be taken with a grain of salt.}
Even without the repulsive interaction, a boson star would form due to the quantum pressure above the line labeled ``Quantum Pressure,'' where eqs.~\eqref{eq:k-Mf} and~\eqref{eq:kheavy} are incompatible.

PBHs form in the region shown in various shades of blue,
given that cannibal reactions are switched off.
This region is magnified in the bottom panels. 
For every point in this region, PBHs are produced within a certain mass range.
The maximum mass~$M_+$ is set by the requirement of a collapse before reheating, as given in eq.~\eqref{eq:Ml_pes}.
The minimum mass~$M_-$ is set either by the requirement of $k < (aH)_i$ in eq.~\eqref{eq:Ms_pes}, or negligible 
repulsive self-interaction in eq.~\eqref{eq:Mcom}, whichever is stronger.
Values of $M_-$ and $M_+$ are shown by the gray dot-dashed and dashed contours, respectively.
(These contours are extended beyond the PBH-forming region for visibility).
As one moves towards the bottom left in the plots,
the PBHs tend to have larger masses and thus longer lifetimes.
We also note that the PBH-forming region vanishes for $\lambda \lesssim 10^{-2}$.

The white-dashed contours show the PBH abundance relative to dark matter, $f_{\rm BH}$, in base 10 logarithm. 
Since $f_{\rm BH}$ scales as~$M_{\rm BH}^{-2.5}$ (see eq.~\eqref{eq:fbh_pes}) and thus the total abundance is dominated by the lightest PBHs, the $f_{\rm BH}$ contours are shown for the minimum mass~$M_-$.
The dark blue region shows where $f_{\mathrm{BH}}>1$; 
in this region one sees that the minimum mass~$M_{-}$ is larger than $10^{15}\, \mathrm{g}$, hence all PBHs survive until today and overdominate the universe. 
On the other hand, in the region with
$M_{-} \lesssim 10^{15}\, \mathrm{g}$,
which appears for $\lambda = 1$,
at least part of the PBHs evaporate 
and are subject to further observational constraints. 
The region labeled ``CMB+BBN'' shows where the PBH spectrum intersects the constraints from CMB and/or BBN (see fig.~\ref{fig:spectrum}) and is therefore excluded. 

The light blue regions show where PBHs are produced with a sufficiently small amount such that they do not overdominate the universe nor violate the CMB/BBN constraints. 
PBHs in most of this region lie in the asteroid-mass window,
and hence also avoid constraints from microlensing observations \cite{Griest:2013esa,Griest:2013aaa,Smyth:2019whb}.
On the left edge of the light blue region,
the PBHs make up the entire dark matter abundance
(i.e. $f_{\ro{BH}} = 1$),
without violating observational constraints.

Note that, for simplicity, we have imposed the conditions such as those described by the ``No Coll'' and ``B$\star$'' lines as hard cutoffs to the PBH formation. However, the actual transition between different regimes is continuous. Thus, at the upper and left edges of the dark blue region, $f_{\ro{NL}}$ should smoothly vary from values larger than unity to zero.
There, the PBHs can also make up the entire dark matter, although a fine-tuning of the parameters would be required.

\subsection{Maximal gravothermal accretion}\label{sec:spec_param}

We now assume maximal gravothermal accretion, i.e. $\eta=\eta_{\rm max}=10^{-3}$ (see eq.~\eqref{eq:opt}). With a sufficient accretion, cannibal and boson stars can eventually collapse into PBHs, which significantly extends the PBH-forming window.

\begin{figure*}
\begin{subfigure}{0.45\textwidth}
    \includegraphics[width=1.00\textwidth]{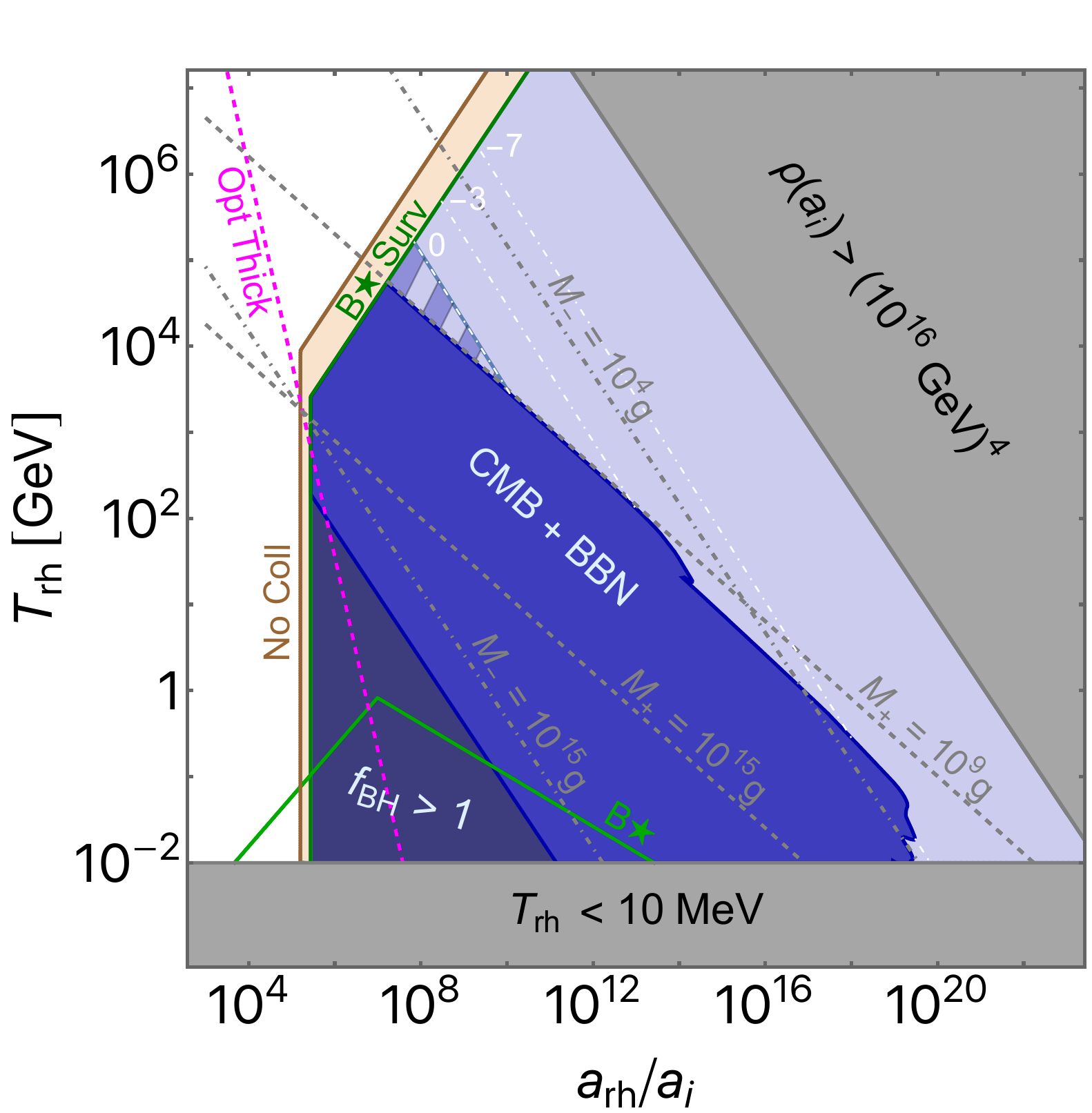}
    \caption{$\lambda = 1,\ \eta=\eta_{\rm max}$, without cannibalism.}
    \label{fig:PBH_3opt}
\end{subfigure}
\begin{subfigure}{0.45\textwidth}
    \includegraphics[width=1.00\textwidth]{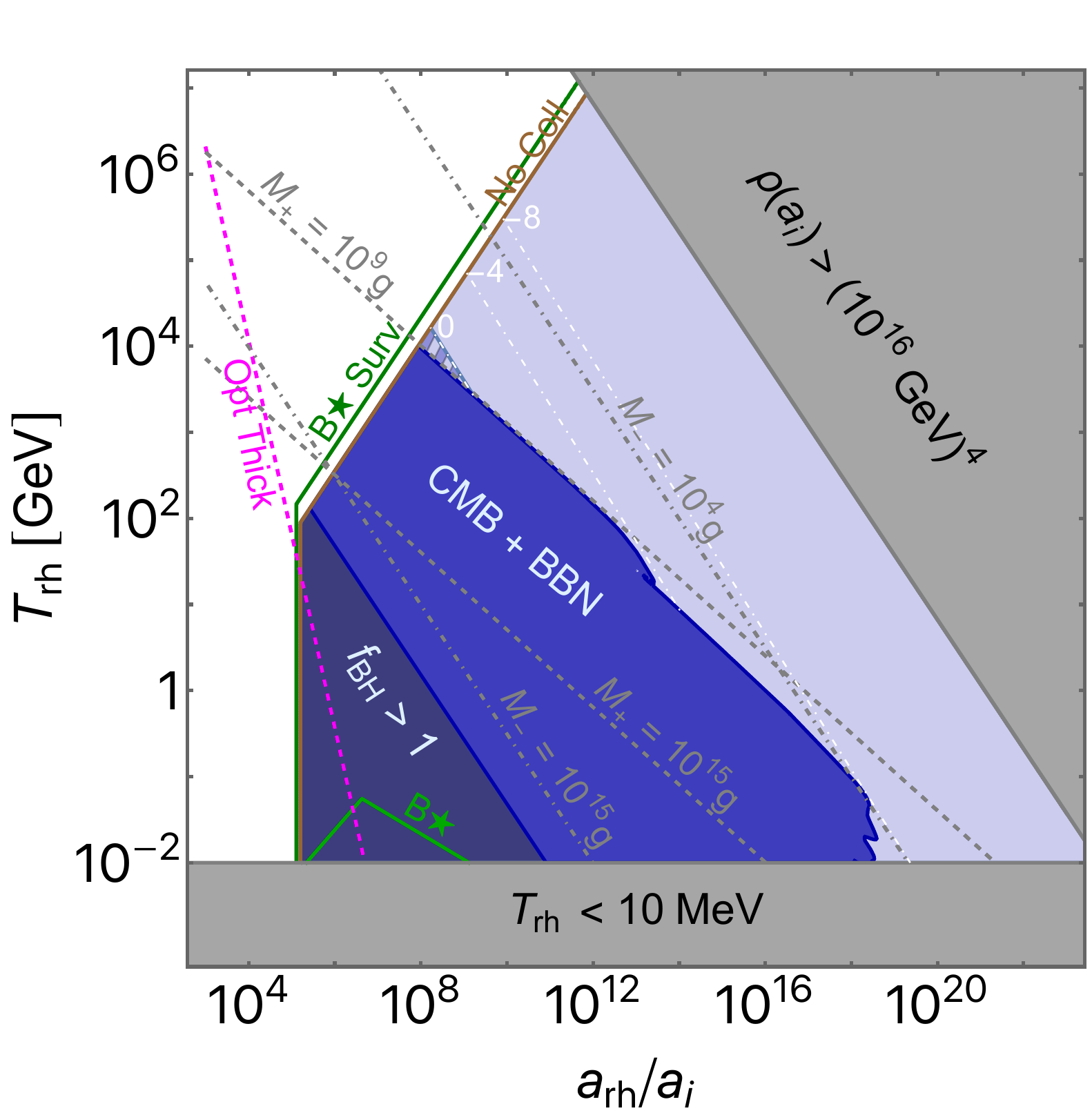}
    \caption{$\lambda = 10^{-1},\ \eta=\eta_{\rm max}$, without cannibalism.}
    \label{fig:PBH_5opt}
\end{subfigure}

\begin{subfigure}{0.45\textwidth}
    \includegraphics[width=1.00\textwidth]{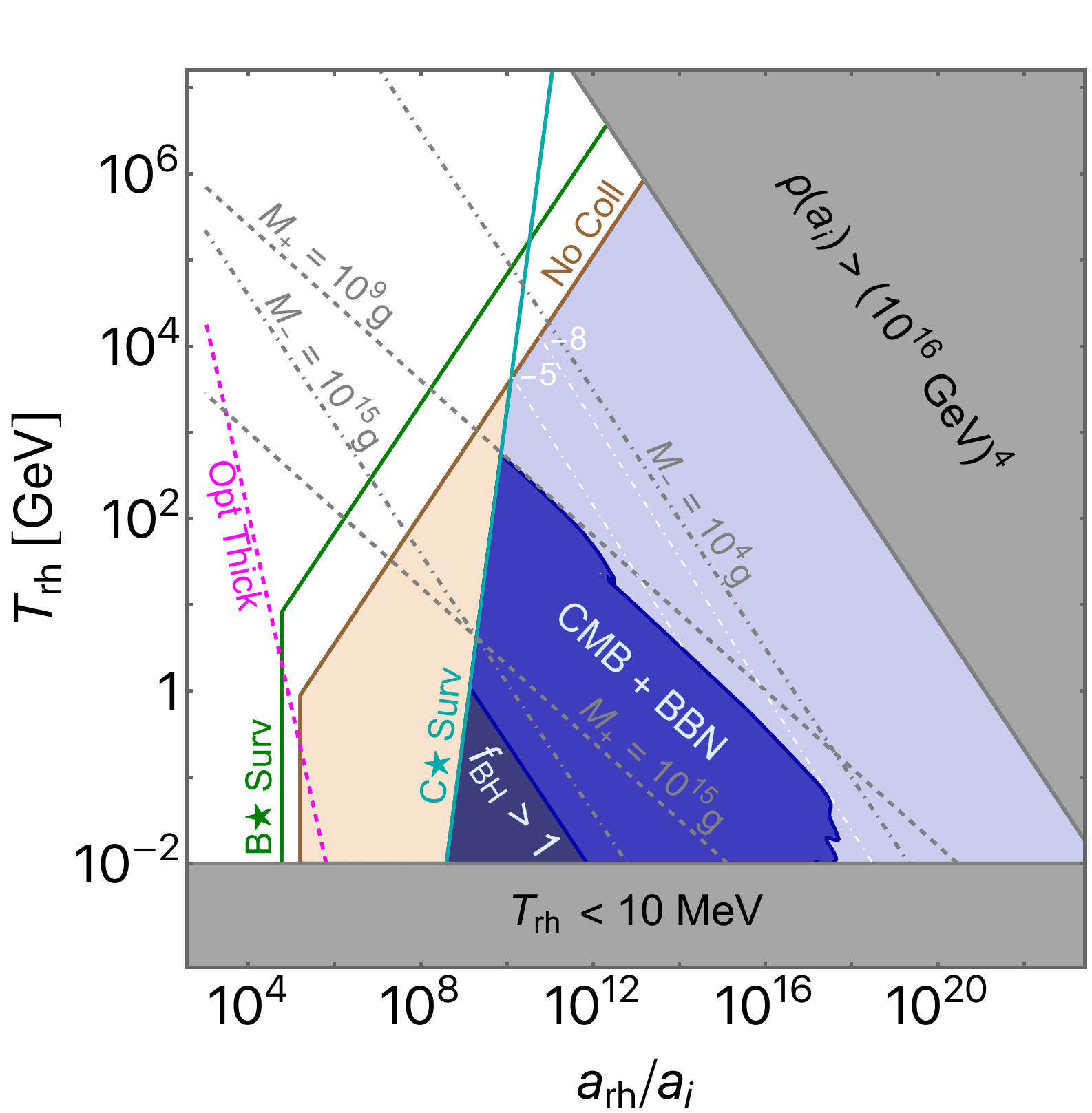}
    \caption{$\lambda = 10^{-2},\ \eta=\eta_{\rm max}$, with cannibalism.}
    \label{fig:PBH_abundance1opt}
\end{subfigure}
\begin{subfigure}{0.45\textwidth}
    \includegraphics[width=1.00\textwidth]{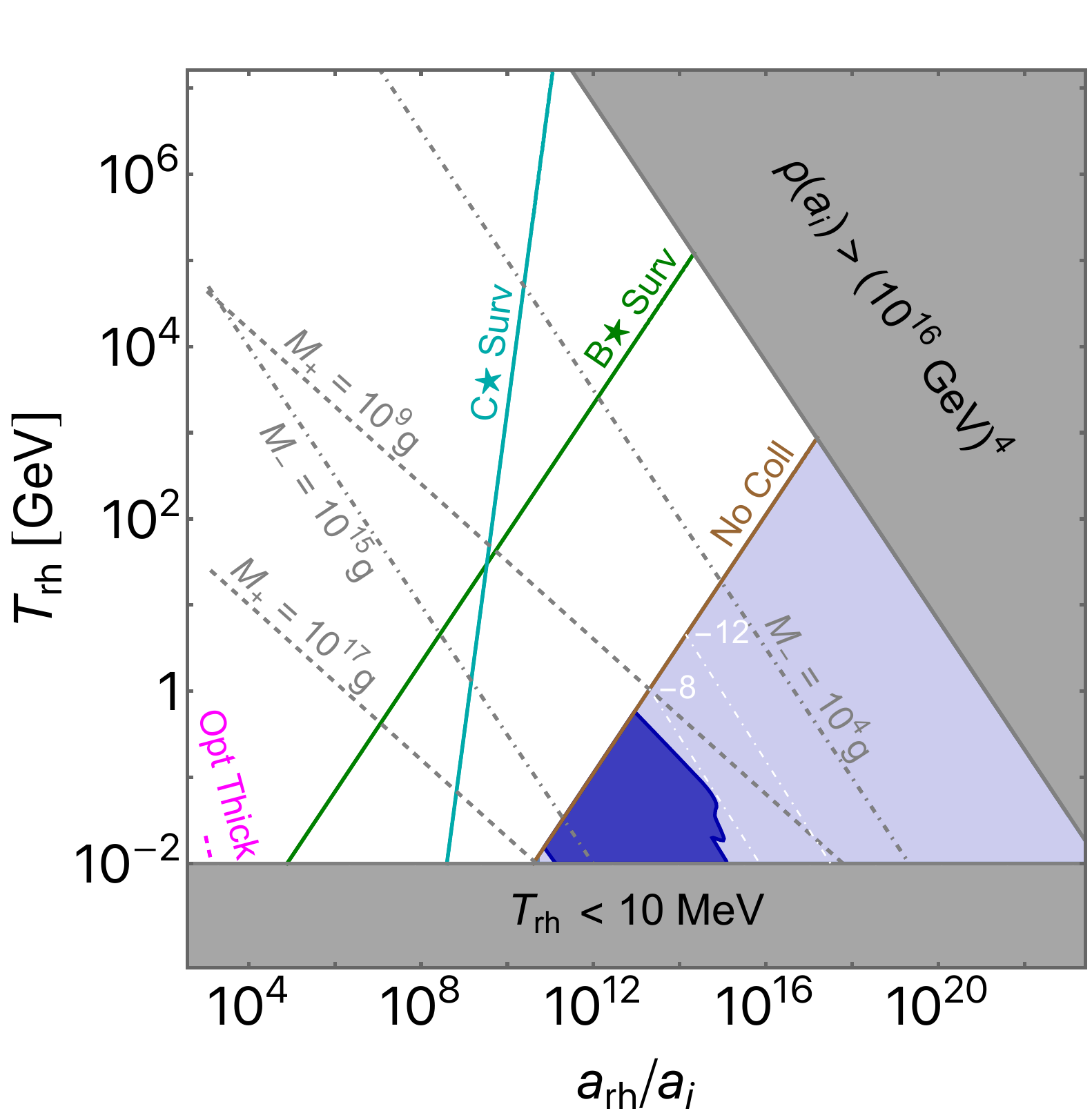}
    \caption{$\lambda = 10^{-5},\ \eta=\eta_{\rm max}$, with cannibalism.}
    \label{fig:PBH_abundance4opt}
\end{subfigure}
\caption{Consequences of gravothermal catastrophe with maximal gravothermal accretion. 
The meanings of the labels and colours are the same as in fig.~\ref{fig:nocan_abundance}, except for that the 
white dot-dashed contours here show the 
PBH abundance relative to the SM particles upon evaporation, $\log_{10} \kappa_{\ro{BH}}$. 
\textbf{Top Panels}: 
Parameter space assuming negligible cannibal annihilations. In the region shown in various shades of blue, PBHs are produced either 
from the direct gravothermal collapse of halos or from the collapse of boson stars. 
In the striped blue region, PBHs come to dominate the universe and evaporate before BBN. 
On the left of the ``B$\star$ Surv'' line, none of the boson stars collapse into PBHs. 
\textbf{Bottom Panels}: 
Parameter space in the presence of cannibal annihilations. In the blue-shaded regions, PBHs are also formed from the collapse of cannibal stars. On the left of the ``C$\star$ Surv'' line, none of the cannibal stars collapse into PBHs.}
\label{fig:PBH_abundance}
\end{figure*}

The top panels of fig.~\ref{fig:PBH_abundance} show parameter spaces with maximal accretion
for $\lambda = 1$ (left) and $10^{-1}$ (right), assuming that cannibal annihilations are switched off. Apart from the value of~$\eta$, the setups are the same as in 
fig.~\ref{fig:nocan_abundance}; thus the positions of the lines for 
``No Coll,'' ``Opt Thick,'' and ``B$\star$'' are also the same. 
(This is the case also for the ``Quantum Pressure'' line, which 
is not shown here for better visibility.)
On the other hand, the PBH-forming region shown in various shades of blue, 
now extends beyond the ``B$\star$'' line.
Here, halo cores first form boson stars, which further collapse into PBHs if the accretion allows them to exceed the critical mass in eq.~(\ref{eq:86}). 
The condition for boson stars to collapse translates into the upper limit on~$k$ in eq.~(\ref{eq:kcom2}), which becomes incompatible with 
eq.~\eqref{eq:kheavy} in the region above the line labeled ``B$\star$ Surv.''
In the orange region in fig.~\ref{fig:PBH_3opt} which is 
sandwiched between the ``B$\star$ Surv'' and ``No Coll'' lines, 
all of the boson stars survive until reheating without collapsing into PBHs.

The bottom panels of fig.~\ref{fig:PBH_abundance} are also for
maximal accretion, but with cannibal annihilations included.
$\lambda $ is taken as $10^{-2}$ (left) and $10^{-5}$ (right);
for these values the condition in eq.~(\ref{eq:76}) for avoiding a cosmological cannibal phase is satisfied in the entire displayed region.
The collapsing halos, on the other hand, are generically affected by the cannibal reactions,  as we discussed around eq.~(\ref{eq:boundary_can}).
In other words, the cannibal star equivalent of the ``B$\star$'' line lies below the plotted region
(as is the case for ``B$\star$'' itself in the lower panels.)
The resulting cannibal stars can collapse into PBHs by growing beyond the threshold mass in eq.~(\ref{eq:73}), or equivalently, for $k$~modes below the upper limit in eq.~(\ref{eq:kcan2}).
This $k$-limit becomes incompatible with eq.~\eqref{eq:kheavy} on the left of the line labeled ``C$\star$ Surv,'' where none of the cannibal stars collapse.

We compute the mass and abundance of PBHs formed by the collapse of stars, in a similar fashion as for those from a direct collapse of halos, by using $ M_{\mathrm{BH}} = \eta_{\ro{max}} M_{\rm halo}$,
and eqs.~(\ref{eq:fbh_opt}) and (\ref{eq:betaopt}). 
Among all PBHs formed in different ways, the maximum mass~$M_+$ is still set solely by the reheating bound in eq.~\eqref{eq:Ml}.
On the other hand, the minimum mass~$M_-$ 
is set by the strongest requirement among
$k < (aH)_i$ in eq.~\eqref{eq:Ms},
boson star collapse in eq.~(\ref{eq:86}),
and if cannibalism is switched on, also by cannibal star collapse in eq.~(\ref{eq:73}).
(Hence in regions where boson and cannibal stars can coexist, we allow PBHs to form if both kinds of stars can collapse.)

With accretion, PBHs form for a wide range of EMDE parameters in both cases with and without cannibalism, and even with $\lambda$ as small as $10^{-5}$. 
The rather efficient PBH production, however, leads to PBH overabundance or exclusion by CMB/BBN limits in a sizeable parameter region. 
The upper right edge of the ``CMB+BBN'' exclusion region 
more or less overlaps with the $M_+ = 10^9\, \ro{g}$ contour.
Thus, the produced PBHs are so abundant that almost all PBHs that evaporate after the onset of BBN are ruled out. 
In figs.~\ref{fig:PBH_3opt}--\ref{fig:PBH_abundance1opt},
the lower left part of the ``CMB+BBN'' region
overlaps with ``$f_{\ro{BH}} > 1$,'' with the former displayed in front of the latter.
Hence, there is no place where the PBHs can explain the entire dark matter while being consistent with observations (except for with fine-tuned parameters on the left edge of the ``$f_{\ro{BH}} > 1$'' region).

Inside the light blue region
where the PBHs are compatible with observations, 
the white dot-dashed contours in fig.~\ref{fig:PBH_abundance} 
show the PBH abundance~$\kappa_{\ro{BH}}$ relative to the SM particles upon evaporation (instead of $f_{\ro{BH}}$).
With maximal accretion, $\kappa_{\ro{BH}}$ increases with the PBH mass  (cf. eq.~\eqref{eq:betaopt}). The $\kappa_{\ro{BH}}$~contours are thus evaluated for~$M_+$ and are shown in regions where the largest PBHs evaporate deep in the radiation-dominated era.
On the other hand, in the upper right part of the light blue region, the entire PBH spectrum 
evaporates before (or upon) reheating.
Here, the density of PBHs until they evaporate is necessarily much smaller than the total density of the universe since only a tiny fraction of the $\phi$ particles end up inside PBHs.

The striped blue region highlights the parameter space where both $\kappa_{\rm BH}>1$ and $M_+ < 10^9\, \ro{g}$ are realized.
Here, the PBHs temporarily dominate the universe, then evaporate before BBN.

\section{Alternative EMDE models}\label{sec:beyond}

We have considered the EMDE to begin when a thermally distributed $\phi$ species becomes non-relativistic. 
However, there can be alternative scenarios.
For instance, an EMDE can arise from a coherent scalar field condensate, such as the inflaton, as it oscillates along a quadratic potential \cite{Turner:1983he, Kane:2015jia}. 
One can also imagine a multi-component universe that is initially dominated by relativistic species, subsequently entering an EMDE as non-relativistic species take over \cite{Zhang:2015era,Ganjoo:2022rhk}.
Furthermore, a cannibal phase could precede the EMDE (cf. discussions in Section~\ref{sec:cosmo_cann}).

Each of the above EMDE scenarios gives a different relation between the particle mass~$m$ and the initial EMDE density~$\rho_\phi (a_{\ro{i}})$.
In our analyses, this amounts to modifying the relation in eq.~(\ref{eq:rho_ai}), which in turn affects 
the relation between the cosmological scales and 
the rates of self-scattering (cf. eq.~\eqref{eq:sigma}), cannibal annihilations (eq.~\eqref{eq:cann}), as well as the boson star mass threshold (eq.~\eqref{eq:Mcom}). 
However, we have checked that with different 
$m$-$\rho_\phi (a_{\ro{i}})$ relations, 
the qualitative picture roughly remains the same, in the sense that
if PBHs heavier than $10^{9}\, \ro{g}$ form, they tend to conflict with observational constraints. 

In scenarios where an EMDE is triggered by species that have always been non-relativistic, perturbation modes that enter the horizon before the onset of EMDE (i.e. $k > (a H)_{\ro{i}}$) can also contribute to the production of collapsing halos.
This would give rise to even smaller PBHs, with which we expect the total PBH abundance to increase 
and thus the constraints become more severe.

One may also consider models with different forms of self-interactions, e.g., cubic or mediated by other particles.
This would significantly modify the relation between the self-scattering and cannibal annihilation rates.
Some models may even prohibit cannibalism, as discussed below 
eq.~(\ref{eq:76}).

\section{Summary and conclusion}
\label{sec:end}

In this work, we studied the gravothermal catastrophe in an EMDE.
We showed that halos formed during an EMDE can undergo a gravothermal collapse for a wide range of self-interaction couplings.
Additionally, we have explored the rich phenomenology that follows from the collapse and showed that it can produce not only PBHs but also other compact objects, such as boson stars and cannibal stars.

The formation of cannibal stars is a particularly salient feature of the gravothermal catastrophe since the self-interactions inducing the thermalization also enable the number-changing interactions that produce heat to support the stars.
For the toy EMDE model with a $\phi^4$~interaction, we found that the gravothermal catastrophe predominantly forms cannibal stars. If the stars accrete the surrounding particles, they can eventually collapse into PBHs. 
Alternatively, in extended models where the number-changing interactions are forbidden, PBHs can form from the direct gravothermal collapse of halos. 

Halos formed during an EMDE have relatively small masses ($M_{\rm halo} \lesssim 10^{28} \, \ro{g}$), resulting in even smaller PBHs following gravothermal collapse. Our toy model reveals that EMDE can produce PBHs across a spectrum of abundances with observational significance: They may be overproduced and conflict with observational constraints; they may form asteroid-mass black holes comprising the entire dark matter content without violating observational limits; or they may temporarily dominate the universe before evaporating prior to BBN.

This study also highlights the potential of using PBHs as a probe of an EMDE. 
For the toy $\phi^4$ model, we found that a rather wide parameter space can be excluded by the overproduction of PBHs. 
It would be interesting to extend our study to other 
well-motivated models, such as those with a dark confining Yang-Mills sector, for which PBH formation may provide novel constraints.

In order to obtain a more accurate prediction of the PBH abundance, our calculations can be improved in the following ways. 
Firstly, it will be important to quantify the gravothermal accretion of the produced objects. 
We have made initial progress by estimating the maximum mass that can be accreted onto black holes from energy conservation considerations. However, a precise estimate of the final PBH mass requires a more detailed analysis. 
We also remark that we have neglected variances in halo properties, 
such as those of the concentration upon halo formation and of the halo mass assembly history.
Moreover, we only considered black hole formation in halos that do not fall into a bigger one until the gravothermal collapse.
Going beyond these approximations can also affect the PBH spectrum.
In particular, the gravothermal collapse of subhalos may significantly enhance the PBH abundance compared to our estimate. 

If subhalos form PBHs and sink toward the host center, it may lead to the formation of PBH binaries, which in turn may produce detectable gravitational wave signals. 
Moreover, various observable signatures arise in the case where an EMDE generates evaporating PBHs, particularly if the PBHs dominate the universe prior to BBN. 
These include
dark matter/radiation \cite{Allahverdi:2017sks,Masina:2020xhk}, 
relic Planck-scale objects \cite{MacGibbon:1987my,Barrow:1992hq,Chen:2002tu},
gravitational waves \cite{Papanikolaou:2020qtd,Papanikolaou:2022chm},
and effects on baryogenesis~\cite{Calabrese:2023key}.
These possibilities highlight the exciting opportunities for future research.

Finally, we remark that while our study focused on PBH formation during EMDE, the rich phenomenology we have found to emerge from the gravothermal catastrophe can have broader applications. For instance, it would be interesting to explore cannibal and boson star formation in the present-day universe through the collapse of self-interacting dark matter halos. 
On a more speculative note, studying star formation and accretion in 
simple particle models, such as the $\phi^4$~model, may provide useful insights into the complex astrophysical processes in our universe. 
We leave an investigation of such possibilities for future work.

\begin{acknowledgments}
We are thankful to Yiming Zhong, Wei-Xiang Feng, Petr Tinyakov, Tomas Reis, Aleksandr Azatov, Kiyotomo Ichiki, Donghui Jeong, Pierluigi Monaco, Aseem Paranjape, Mak Pavi\v{c}evi\'{c}, Serguey Petcov, and Emiliano Sefusatti for useful discussions.
This research was supported in part by INFN TASP.
T.K. also acknowledges support from the European Union - NextGenerationEU through the PRIN Project ``Charting unexplored avenues in Dark Matter'' (20224JR28W), and from JSPS KAKENHI (JP22K03595). 
D.P. was partially supported by the National Science Centre,
Poland, under research
grant no. 2020/38/E/ST2/00243.
\end{acknowledgments}

\appendix

\section{Density perturbation evolution during matter domination}\label{app:perturb}
In this appendix, we review the evolution of density perturbations during a matter-dominated era.

We work in conformal Newtonian gauge with metric given by
\begin{align}
ds^2=-(1+2\psi)dt^2+a^2(t)(1-2\Phi)d\bd{x}^2,
\end{align}
where $\Phi$ and $\psi$ are scalar metric perturbations. As the anisotropic stress of non-relativistic matter is negligible, we have $\psi=\Phi$.

We are interested in the evolution of the matter density perturbation, $\delta_m=(\rho_m - \bar{\rho}_m) / \bar{\rho}_m$, where $\bar{\rho}_m$ is the background matter density. Deep inside the horizon, $\delta_m$
and the metric perturbation are related to each other by the Poisson's equation. This equation in Fourier space is given by
\begin{align}
    k^2\Phi(k)=-4\pi G\bar{\rho}_m a^2 \delta_m(k,a).
\end{align}
Using the fact that the Hubble rate is determined by $\bar{\rho}_m$, the above can be rewritten as
\begin{align}
    \delta_m(k,a)=-\frac{2}{3}\frac{k^2}{(aH)^2}\Phi(k).
\end{align}

An interesting feature of the conformal Newtonian gauge is that during matter domination, $\Phi(k)$ remains constant even through horizon crossing. 
Moreover, while the mode~$k$ is outside the horizon, 
$\Phi(k)$ is related to the gauge-invariant primordial curvature perturbation $\mathcal{R}(k)$ via $\Phi(k)=-3\mathcal{R}(k)/5$.
Hence, we can connect the density perturbation inside the horizon to the curvature perturbation before the mode entered the horizon as
\begin{align}
    \delta_m(k,a)=\frac{2}{5}\frac{k^2}{(aH)^2}\mathcal{R}(k).
\end{align}
Writing the scale factor at horizon entry of the mode as $k\equiv (aH)_{\mathrm{hor}}$, and using the fact that $H\propto a^{-3/2}$,
the above simplifies to
\begin{align}
    \delta_m(k,a)=\frac{2}{5}\frac{a}{a_{\mathrm{hor}}}\mathcal{R}(k).
\end{align}
This expression describes the growth of linear density perturbations inside the horizon for modes that enter the horizon during matter domination.

\section{Evolution of halos using Press--Schechter formalism}\label{app:press}

The Press--Schechter formalism is a key tool in cosmology for predicting the abundance of halos based on their mass. The formalism only takes in the linear matter power spectrum as input to yield the halo mass function,
\begin{equation} \label{eq:HMFdndM}
    \frac{dn_{\rm h}}{dM} (M,a) = \sqrt{\frac{2}{\pi}}\frac{\Bar{\rho}_{\rm m}}{M} \frac{\delta_c}{\sigma_{\rm R}^2} 
\left| \frac{d  \sigma_{\rm R}}{dM} \right|
e^{-\delta_c^2/[2\sigma_{\rm R}^2]},
\end{equation}
where $n_{\rm h}$ is the number density of halos up to mass $M$, 
and $\delta_c=1.686$ is the critical threshold beyond which linear density perturbations collapse to form halos. Moreover, $\sigma^2_{\rm R}$ is the variance of the smoothed linear overdensity field of the total matter and is determined by
\begin{equation}\label{eq:sigma_halo}
    \sigma^2_{\rm R} \equiv \langle \delta^2 (\bd{x},a;{\rm R})\rangle =  \int_0^{\infty} 
\frac{d k}{k}  \Delta^2_\delta(k,a)
\Tilde{W}^2(k{\rm R})  .
\end{equation}
Here $\Delta^2_\delta$ is the dimensionless power spectrum of the linear density perturbation, and 
$\Tilde{W}$ is a Fourier-transformed window function, with the smoothening scale given by the radius ${\rm R} = (3M/4\pi \Bar{\rho}_{\rm m})^{1/3}/a$. (Note that $\mathrm{R}$ is time independent for a fixed~$M$ since $\bar{\rho}_{\ro{m}} \propto 1/ a^{3}$.)

The number density of halos of a given mass, $M$, goes through two distinct phases of evolution. The first phase is while $\sigma_{\rm R}\ll \delta_c$, or equivalently, when the density perturbations corresponding to the scale R are still linear. In this phase, the exponential term in eq.~\eqref{eq:HMFdndM} dominates, and the abundance of halos exhibits an exponential rise with time.

The exponential rise continues until $\sigma_{\rm R}\sim \delta_c$. Around this point, the abundance reaches its maximum such that almost all the matter is collapsed in halos of mass $M$,
\begin{equation}
\begin{split}
 \frac{d\rho_{\rm h}}{d\ln M} 
&=M\frac{d n_{\rm h}}{d\ln M} \\
& = \sqrt{\frac{2}{\pi}}\Bar{\rho}_{\rm m} \frac{\delta_c}{\sigma_{\rm R}} 
\left| \frac{d\ln  \sigma_{\rm R}}{d\ln M} \right|
e^{-\delta_c^2/[2\sigma_{\rm R}^2]}
\sim \Bar{\rho}_{\rm m}.
\end{split}
\end{equation}

As the scale $\ro{R}$ goes further into the non-linear regime, i.e. $\sigma_{\rm R}\gg \delta_c$, the number of halos of mass $M$ decreases because they are accreted into larger halos. One can find the rate at which the abundance falls by approximating the exponential in eq.~\eqref{eq:HMFdndM} as unity to obtain
\begin{align}
    \frac{d n_{\rm h}}{d M} (M,a) \approx \sqrt{\frac{2}{\pi}}\frac{\Bar{\rho}_{\rm m}}{M} \frac{\delta_c}{\sigma_{\rm R}^2} 
\left| \frac{d  \sigma_{\rm R}}{d M} \right| .
\end{align}
There are only two terms on the right-hand side with time dependence: $\bar{\rho}_m$ and $\sigma_{\rm R}$. The matter density simply evolves as $\bar{\rho}_m\propto 1/a^3$. Since during the matter-dominated era, the matter density perturbations grow proportional to the scale factor
(cf. eq.~\eqref{eq:delta_ev}), we find that $\sigma_{\rm R}\propto a$. Thus,
$dn_{\ro{h}}/dM \propto 1/a^{4}$, or equivalently, the comoving number density of halos\footnote{Crudely, one considers the increase in time of 
$dn_{\ro{h}}/dM$ due to the factor $\exp (- \delta_c^2 / 2 \sigma_{\ro{R}}^2 )$ to correspond to the formation of halos, and the decrease due to $\delta_c / \sigma_{\ro{R}}$ as accretion into larger halos
(here $d \ln \sigma_{\ro{R}} / d M$ is considered to be constant).
This suggests that most of the halo formation takes place during the linear regime, i.e., $ \sigma_{\ro{R}} \lesssim \delta_c $.\label{foot:26}}
falls as $1/a$.

\bibliography{apssamp}

\end{document}